\documentclass[floatfix,10pt,aps,prb,twocolumn,showpacs,a4paper, jpconf]{revtex4}
\usepackage{amsmath}   % contains the advanced math extensions for LaTeX
\usepackage{mathtools} % an extension package to amsmath
\usepackage{amssymb}   % adds new symbols in to be used in math mode
\usepackage{mathrsfs}  % other mathematical symbols
\usepackage{textcomp}  % provides extra symbols
\usepackage{accents}   % provides some miscellaneous tools for mathematical accents
\usepackage{bm}        % provides bold font for math symbols
\usepackage{soul}
\usepackage{relsize}
\usepackage{graphicx}  % to manage external pictures
\usepackage{epstopdf}  % adds support of handling eps images to package graphics or graphicx
\usepackage{subfigure} % provides support for the inclusion of small, ‘sub’, figures and tables
\usepackage{makecell}
\usepackage{hhline}
\usepackage{cellspace}
\usepackage{array}     % extends the possibility of LaTeX to create very complicated and customized tables
\usepackage{tabularx}  % modifies the widths of certain columns, rather than the inter column space
\usepackage{multirow}  % provides a construction for table cells that span more than one row of the table
\usepackage{booktabs}  % publication quality tables
\usepackage{dcolumn}   % defines a system for defining columns of entries in an array or tabular which are to be aligned on a 'decimal point'
\usepackage{gensymb}   % for \degree and \celcius
\usepackage{setspace}
\usepackage{scrextend}
\usepackage{multirow}
\newcommand{\angstrom}{\mbox{\normalfont\AA}}

\usepackage[usenames,dvipsnames]{color} % to use color text in LaTeX2e
\usepackage{url}       % defines the \url{...} command
\usepackage{hyperref}  % extends the functionality of all the LaTeX cross-referencing commands
\hypersetup{colorlinks,citecolor=Violet,linkcolor=Red,urlcolor=Blue}

\graphicspath{{./}}
\begin{document}
\setlength{\heavyrulewidth}{0.08em}
\setlength{\lightrulewidth}{0.05em}
\setlength{\cmidrulewidth}{0.03em}
\setlength{\belowrulesep}{0.65ex}
\setlength{\belowbottomsep}{0.00pt}
\setlength{\aboverulesep}{0.40ex}
\setlength{\abovetopsep}{0.00pt}
\setlength{\cmidrulesep}{\doublerulesep}
\setlength{\cmidrulekern}{0.50em}
\setlength{\defaultaddspace}{0.50em}
\setlength{\tabcolsep}{4pt}
\title{\textit{Ab initio} description of magnetic and critical properties of spin-glass pyrochlore NaSrMn$_{2}$F$_{7}$}
\author{Mohammad Amirabbasi} \email{mo.amirabbasi@gmail.com}
%\email{mo.amirabbasi@gmail.com}
\affiliation{Department of Physics, Isfahan University of Technology (IUT), Isfahan 84156-83111, Iran}
\affiliation{Faculty of Physics, University Duisburg-Essen, 47057 Duisburg, Germany}
\affiliation{Corresponding author email: mo.amirabbasi@gmail.com}

%\email{mo.amirabbasi@gmail.com}
\date{\today}
%---------- ABSTRACT -------------------------------------------------------------------------------------------------------------
\begin{abstract}
In this study, I have investigated the magnetic and critical properties of manganese pyrochlore fluoride NaSrMn$_{2}$F$_{7}$, which exhibits a glass transition at T$_\text{f}$$=$2.5 (K) due to charge disorder. 
A DFT+$U$+SOC framework is used in this paper to derive spin-Hamiltonian terms, including isotropic and anisotropic exchange interactions.
An optimized geometry reveals a local distortion of the F-Mn-F angle along the $ < $111$ > $ direction (95.48$^{\circ}$ and 84.51$^{\circ}$), which is considered a weak bond disorder ($\delta J$).
In spite of the complex structure of this material, first principle calculations show that its magnetic properties are only controlled by the nearest neighbor's Heisenberg exchange interaction, and other interactions do not affect spin arrangements in the ground state. Thus, this material is considered a suitable candidate for studying electron correlation in spin glasses.
Using a replica-exchange framework, Monte Carlo simulations indicate that with $\delta J$=0, no phase transition is observed when magnetic susceptibility changes with temperature.
Based on $\delta J$ and the spin Hamiltonian, 2.6 (K) is obtained as the phase transition temperature.
\end{abstract}
%-----------------------------------------------------------------------------------------------------------------------------------
\maketitle

%%%%%###########Introduction#######################################################################################################
\section{introduction}
\label{sec:introduction}
Spin-glass (SG) is a type of magnetic phase in which the spatial arrangement of spins at low-temperatures is completely random~\cite{mydosh1993spin, Binder1986, taniguchi2009spin, shintani2006frustration}.
%%Fig.1%%%%%%%%%%%%%%%%%%%%%%%%%%%%%%%%%%%%%%%%%%%%%%%%%%%%%%%%%%%%%%%%%%%%%%%%%%%%%%%%%%%%%%%%%%%%%%%%%%%%%%%%%%%%%%%%%%%%%
\begin{figure*}[!htp]
    \centering
    \includegraphics[width=0.61\columnwidth]{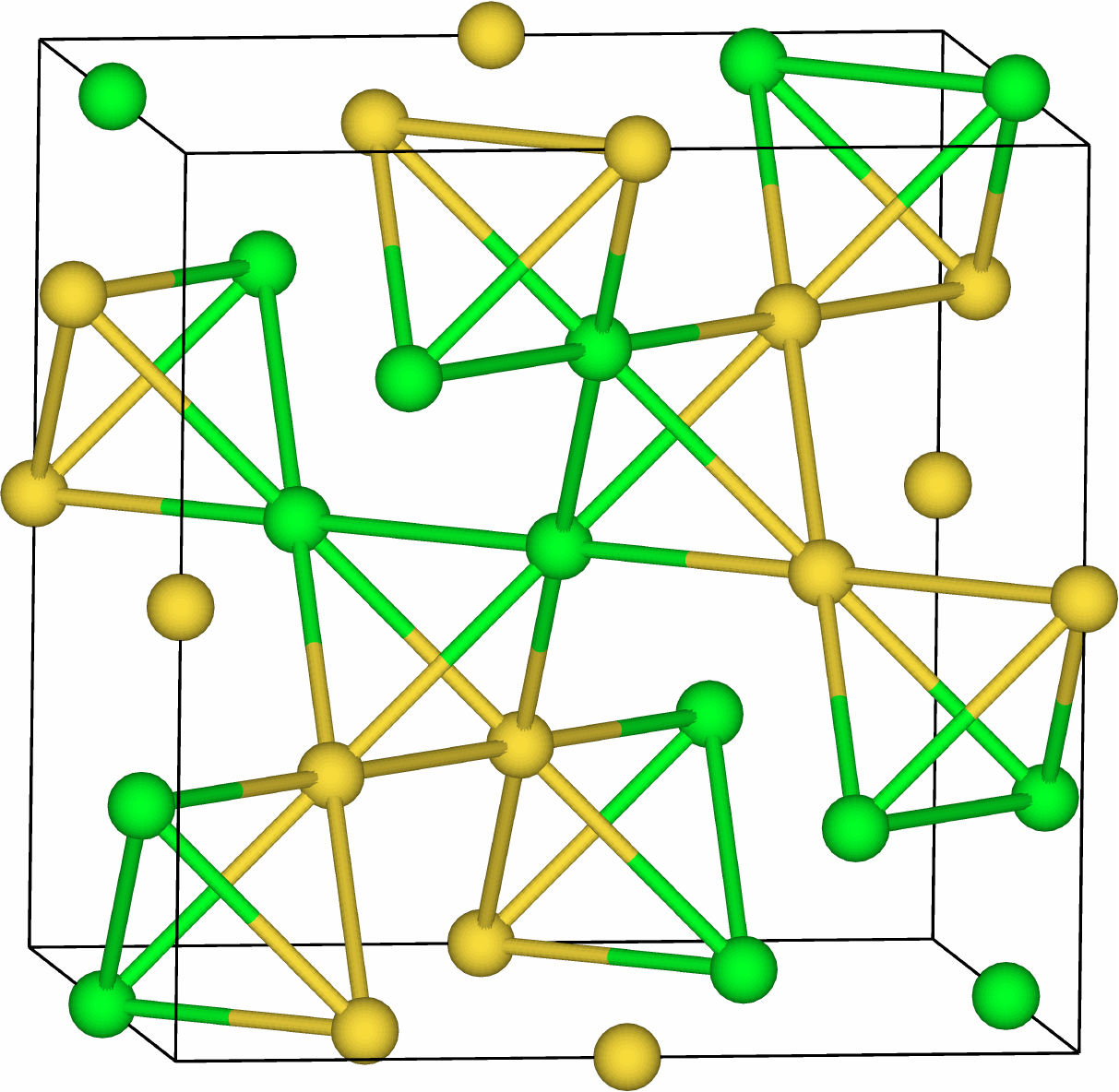}
    \hspace*{1cm}
    \includegraphics[width=0.61\columnwidth]{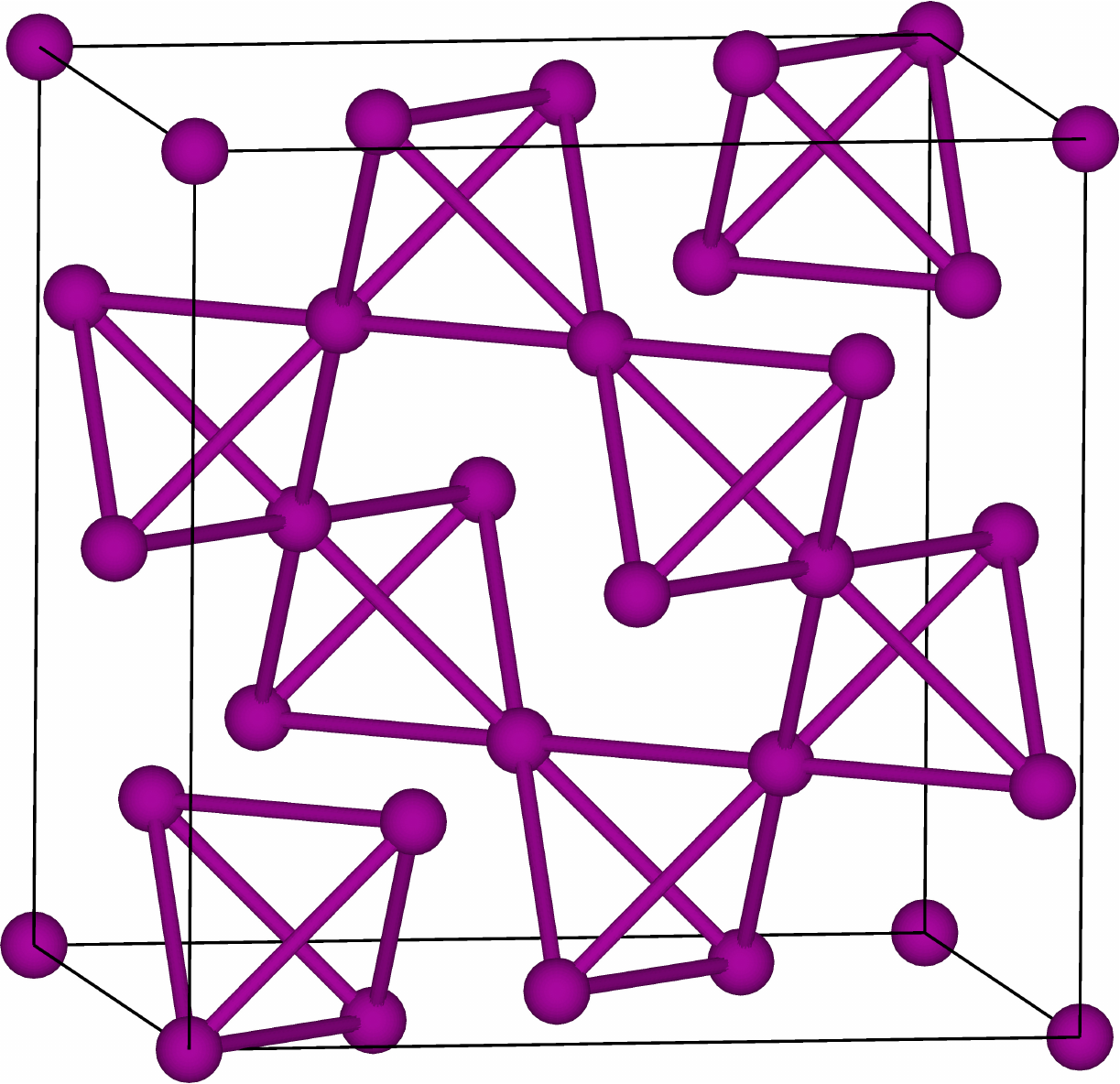}
        \includegraphics[width=0.55\columnwidth]{distortion.pdf}
    \caption{(Color online) Crystal structure of SG pyrochlore NaSrMn$_{2}$F$_{7}$  with corner-shared tetrahedrons for ({\bf Left}) A-site (Na-Sr atoms) and ({\bf right}) B-site (Mn atoms).  Green, yellow and purple spheres are Na$^{1+}$, Sr$^{2+}$ and Mn$^{2+}$, respectively. 
    $x$=0.3331 is experimental value which is reported from the X-ray measurement~\cite{sanders2016nasrmn2f7}.
    The experimental F-Mn-F bond angles in $<$111$>$ direction, lattice constant and F-Mn bond length  in this material are 80.9 and 99.1$^{\circ}$, 2.06 \angstrom and 10.8012(3) \angstrom, respectively~\cite{sanders2016nasrmn2f7}.}
    \label{fig:struct_pyr}
\end{figure*}
    %%%%%%%%%%%%%%%%%%%%%%%%%%%%%%%%%%%%%%%%%%%%%%%%%%%%%%%%%%%%%%%%%%%%%%%%%%%%%%%%%%%%
The presence of frustration (coexisting and competing ferromagnetic and antiferromagnetic interactions)  and randomness (such as chemical disorders or weak bond disorders) are essential for glass formation~\cite{mydosh1993spin, Binder1986, KAWAMURA20151}.
These materials undergo a second-order phase transition which makes the study of SG materials attractive to scientists~\cite{KAWAMURA20151}.
Experimental evidence to detect such a phase are bifurcation of the field cooled (FC) and zero field cooled (ZFC) of DC susceptibility, frequency dependence of AC susceptibility, no magnetic saturation even at low-temperature and high magnetic field and diffusive peaks inelastic neutron scattering pattern~\cite{mydosh1993spin, taniguchi2009spin, vincent2018spin}.
In recent years, some geometrically frustrated pyrochlores have shown the SG phase~\cite{Mitsumoto2020, Shinaoka2011}. 
These materials are composed of a network of corner-sharing tetrahedra~\cite{SUBRAMANIAN198355, Gardner2010}. 
If the exchange interaction between the nearest neighbor is antiferromagnetic, the system fails to reach the ground state with a global minimum.
This condition is called geometric frustration~\cite{SUBRAMANIAN198355, Gardner2010, lacroix2011introduction, Reig-i-Plessis2021}. Despite many studies and researches, there are still many questions about what is the truly thermodynamic SG transition~\cite{Binder1986}.
Determination of the ground state of classical Heisenberg antiferromagnetic SG pyrochlores in which only Heisenberg interactions play a key role can be a good candidate to answer these questions.

In 1997, with the discovery of the spin-ice phase in Ho$_{2}$Ti$_{2}$O$_{7}$~\cite{Harris1997}, study on the magnetic properties of oxide pyrochlores increased~\cite{slobinsky2021monopole, moessner2003theory, marlton2021lattice, yahne2021understanding, shi2021ice, fennell2009magnetic, bramwell2001spin}.
One of the main problems in the study of these materials is the difficult conditions of their crystal growth. 
As a consequence, the availability of large crystal pyrochlores in large sizes has made it impossible.
There is no such problem for  pyrochlore fluoride with general chemical formula A$_{2}$B$_{2}$F$_{7}$ in which Na$^{1+}$/Sr$^{2+}$(Ca$^{2+}$) atoms occupy A-site position randomly and magnetic atoms such as Mn$^{2+}$, Fe$^{2+}$, Co$^{2+}$ and Ni$^{2+}$ occupy B-site~\cite{sanders2016nasrmn2f7, Krizan2014, krizan2015, krizan2015nasrco2f7}.
The result of this random occupation is the presence of the $\delta J$ (difference in the bond angle between the magnetic ions). Therefore, all of these materials exhibit SG properties at low-temperatures~\cite{Shinaoka2011, Andreanov2010}. 
In this paper, I intend to investigate the magnetic and critical properties of NaSrMn$_{2}$F$_{7}$ which has the largest spin moment among the pyrochlore fluoride by first-principle methods in the framework of DFT and Monte Carlo (MC) simulation.
In this material, the Mn$^{2+}$ terminates in 3d$^{5}$.
Therefore, S=5/2 and the orbital moment has a small contribution to the effective moment (6.25 $\mu_{B}$/Mn~\cite{sanders2016nasrmn2f7}). The absence of orbital moment and its associated complexities make this material a suitable candidate to study the effect of bond weak disorder in the phase transition of SG ground state.

The lattice of the NaSrMn$_{2}$F$_{7}$ consists of two separate sublattices with corner-shared tetrahedra (Fig.~\ref{fig:struct_pyr}). In one sublattice, the  Na$^{1+}$/Sr$^{2+}$ ions and in the other the magnetic ion (Mn$^{2+}$) sit on the vertices of each tetrahedra.
The space group of this pyrochlore is Fd-3m (No. 227). Accordingly, the Na$^{1+}$/Sr$^{2+}$, Mn$^{2+}$ and F$^{1-}$ occupy the 16d (0.5, 0.5, 0.5), the 16c (0, 0, 0), 48f ($x$, 1/8, 1/8) and the 8b (3/8, 3/8, 3/8), respectively.
The only variable parameter in this structure is the $x$-positional parameter. 
This parameter ($x$$=$0.3331~\cite{sanders2016nasrmn2f7}) controls the bond angle (Mn-F-Mn) and bond length (Mn-F) which plays a decisive role in the magnetic properties.
Experimental measurements show that this material is an insulator and shows a SG transition at 2.5~(K)~\cite{sanders2016nasrmn2f7}. 
The most important reason for SG transition can be attributed to the change in the bond angle due to the presence of charge disorder (Na$^{1+}$/Sr$^{2+}$) in the lattice~\cite{sanders2016nasrmn2f7}.
The interaction between Mn$^{2+}$ ions is the superexchange mediated by F$^{1-}$.
This mechanism is directly related to the bond length and bond angle. Around each Mn ions, there is a flattened octahedra by F ions. In this octahedra, the F-Ni-F angles are different in $<$111$>$ direction (97.8$^{\circ}$ and 82.2$^{\circ}$~\cite{sanders2016nasrmn2f7}).
The change in F-Ni-F bond angle causes variations in exchange parameters. 
As a result, the ground state of the system takes SG order.
The Curie-Weiss temperature ($\theta_\mathrm{CW}$) equals $-$89.72~(K) which indicates the presence of strong antiferromagnetic interactions between magnetic ions~\cite{sanders2016nasrmn2f7}.

Microscopic examination of the exchange interactions in this material and its effect on the formation of the SG phase as well as the study of the critical properties seem interesting.
This paper aims to determine magnetic and critical properties of NaSrMn$_{2}$F$_{7}$ by constructing a spin Hamiltonian model. It includes various interactions such as the Heisenberg exchange parameters up to the third neighbor, the bi-quadratic, single-ion anisotropy (SIA) and Dzyaloshinskii-Moriya interactions (DMI) as obtained by the first-principle methods. 
By using this information, one can obtain the critical properties of this material, including transition temperature.
The current calculations show that only the $J_{1}$ + $\delta{J}$ is important in determining the magnetic properties of this material, and the effect of other interactions is negligible.
As a result, this material is considered as a simple model of SG and it can be a good candidate for further studies such as understanding the electron-electron correlation influence on magnetic ground state~\cite{alexandradinata2020future} especially in the SG phase, the effect of $\delta{J}$ on thermal fluctuations and removing huge degeneracy~\cite{Andreanov2010} and what is the border between SG and liquid phase according to the strength of $\delta{J}$~\cite{Moessner1998, yang2015spin, cepas2012heterogeneous}.

The paper is structured as follows. In section~\ref{sec:methodology} the details of the DFT and MC computational
methods have been presented.
Section~\ref{sec:results} is devoted to deriving the spin Hamiltonian and the different aspects of exchange interactions in
determination of  magnetic ground state of NaSrMn$_{2}$F$_{7}$ and in the following, the critical properties
are discussed. 
Finally, In section~\ref{sec:conclusion} a summary is given.
%Fig.2%%%%%%%%%%%%%%%%%%%%%%%%%%%%%%%%%%%%%%%%%%%%%%%%%%%%%%%%%%%%%%%%%%%%%%%%%%%%%%%%%%%%%%%%%%%%%%%%%%%%%%%%%%%%%%%%%%%%%%
\begin{figure*}[!htp]
    \centering
    \includegraphics[width=0.5\columnwidth]{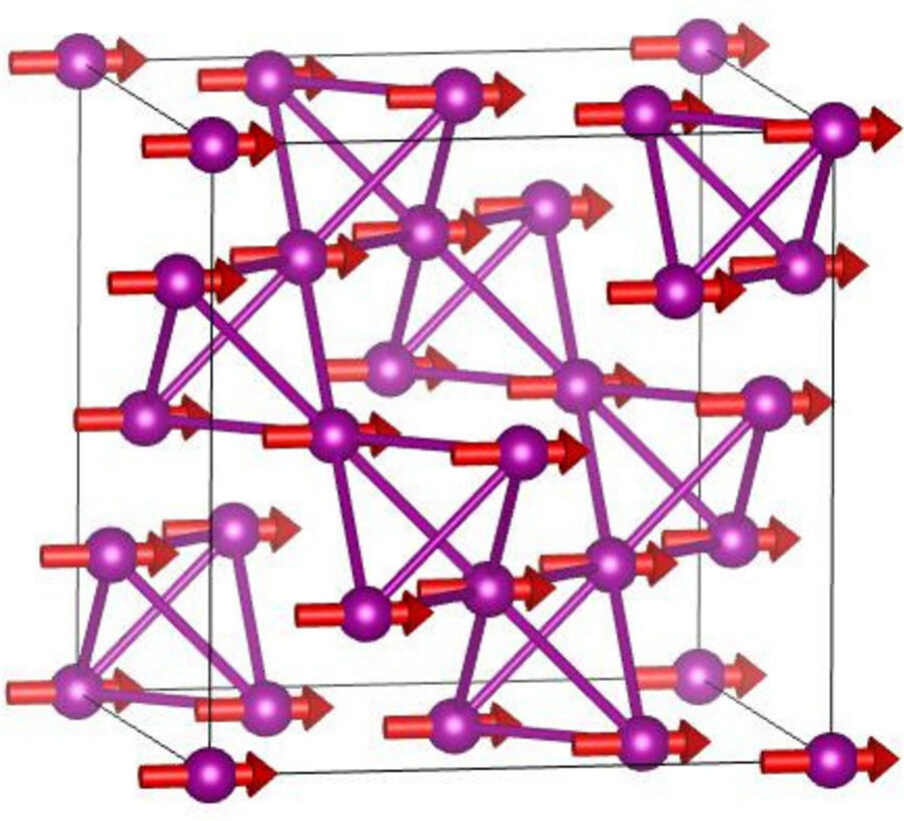}
    \includegraphics[width=0.5\columnwidth]{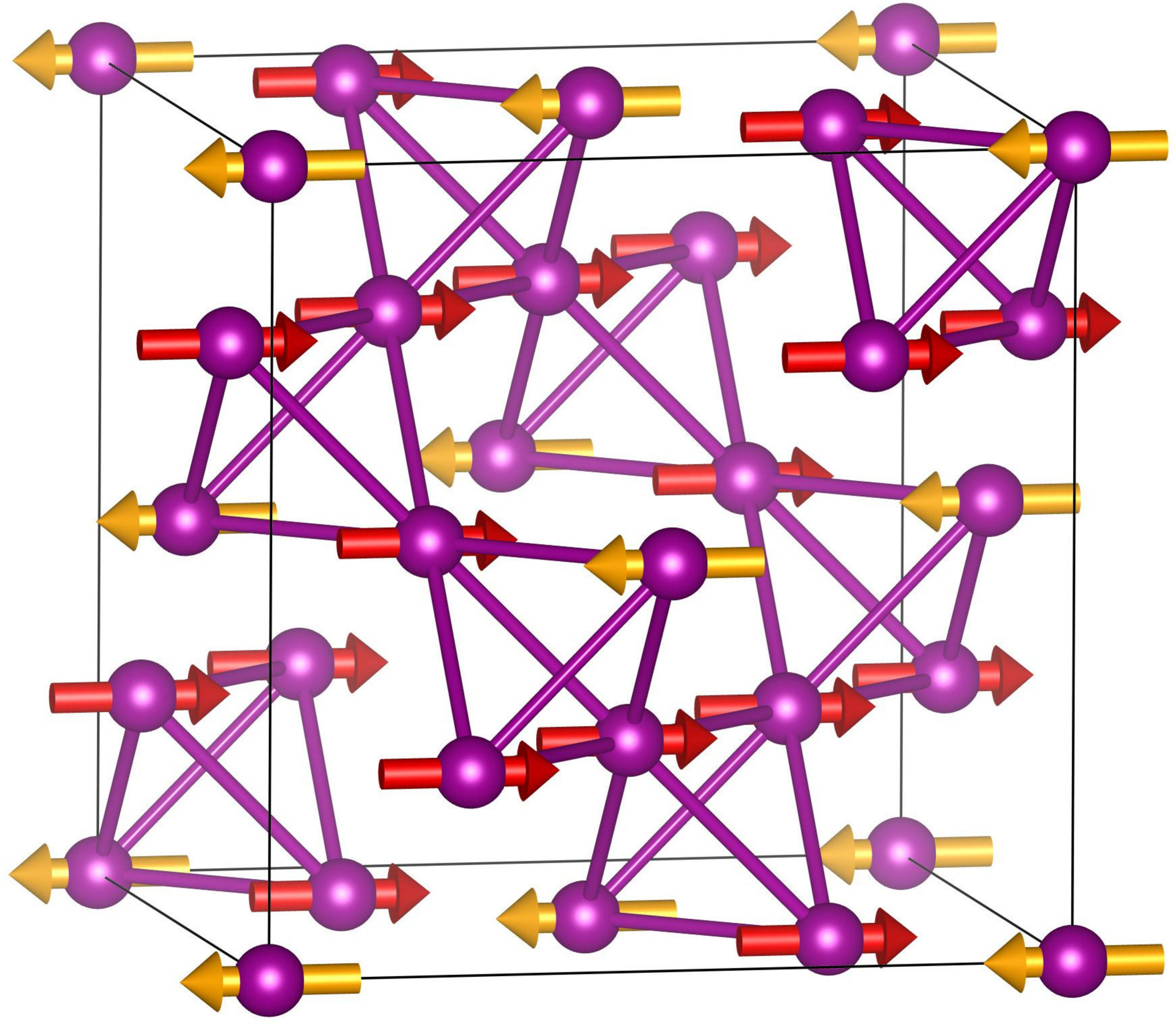}
   %\\*[5mm]
    \includegraphics[width=0.5\columnwidth]{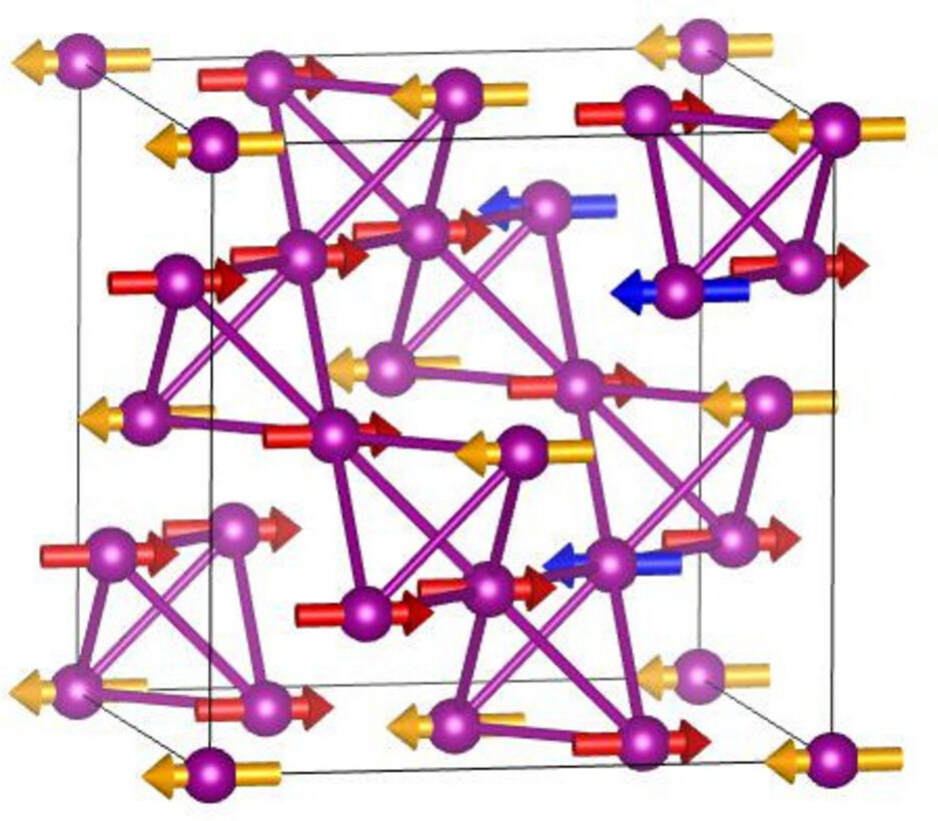}
     %\hspace*{3cm}
    \includegraphics[width=0.5\columnwidth]{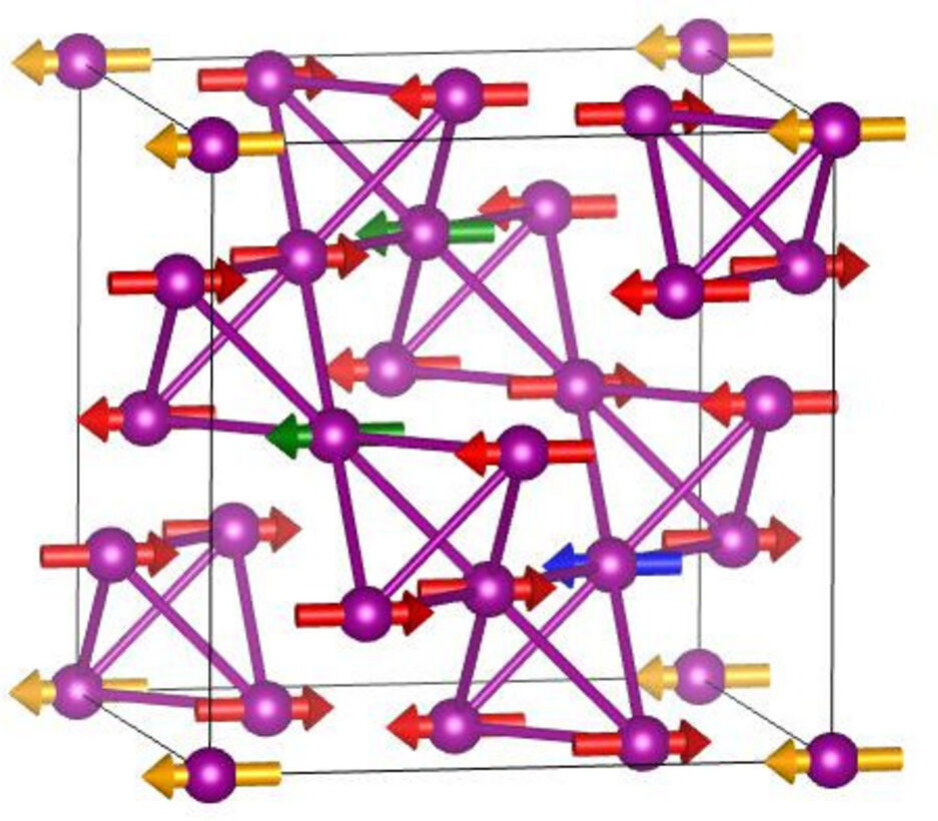}
   \caption{(Color online) Different collinear spin configurations for calculation Heisenberg couplings. The purple spheres denote Mn$^{2+}$. The arrows indicate the direction of the spins for each atom.
   From left to right, the colors of some of the spins are selected differently to distinguish between different configurations. The ferromagnetic configuration has been chosen as the reference to calculate the total energy difference of the various configurations. 
   In this paper, 20 different spin configurations have been used to make sure that the $J_{1}$ converges with respect to the number of configurations.}
   \label{fig:Hzn}
\end{figure*}
%%Computational methods%%%############################################################################################################
\section{Computational methods}
\label{sec:methodology} 
%In this paper, the experimental lattice constant and $x$-positional parameter have been optimized.
In this paper, I introduce the classical spin Hamiltonian ($H$) to get magnetic properties of SG pyrochlore NaSrMn$_{2}$F$_{7}$:
%%%%%%%%%%%%%%%%%%%%%%%%%%%%%%%%%%%%%%%%%%%%%%%%%%%%%%%%%%%Eq.1%%%%%%%%%%%%%%%%%%%%%%%%%%%%%%%%%%%%%%%%%%%%%%%%%%%%%%%%%%%%%%%%%%%%
\begin{equation}
\label{H}
\begin{split}
  H_{\rm {spin}} = - \frac{1}{2}\sum_{i\neq j} J_{ij}(\vec{S_{i}}\cdot\vec{S_{j}})+\frac{1}{2} B\sum_{\rm n.n} (\vec{S_{i}}\cdot\vec{S_{j}})^{2} \\
+\frac{1}{2} D \sum_{\rm n.n} \hat{D}_{ij}\cdot(\vec{S_{i}}\times \vec{S_{j}})+\frac{1}{2} \Delta\sum_{i} (\vec{S_{i}}\cdot\vec{d_{i}})^{2}
\end{split}
\end{equation}
%%%%%%%%%%%%%%%%%%%%%%%%%%%%%%%%%%%%%%%%%%%%%%%%%%%%%%%%%%%%%%%%%%%%%%%%%%%%%%%%%%%%%%%%%%%%%%%%%%%%%%%%%%%%%%%%%%%%%%%%%%%%%%%%%%%%
where $\vec{S_{i}}$ denotes magnetic spins, vector $\hat{D}_{ij}$ shows direction of DM which is determined by Moriya's rules ~\cite{Moriya1960, Elhajal2005} and vector $\vec{d_{i}}$ is the single-ion easy-axis direction at each site $i$.
According to Moriya's rules, if the mirror planes join two sites and the middle point of the opposite bond then the  DM vectors should be perpendicular to the mirror plane~\cite{Moriya1960, Elhajal2005}. It should be noted that in this model $|\vec{S}|$=1.
Different terms of spin Hamiltonian express different types of exchange interactions without (isotropic terms) and with (anisotropic terms) considering spin-orbit coupling (SOC).
Isotropic terms, $J_{ij}$ and $B$ are Heisenberg and bi-quadratic coupling, respectively. Heisenberg interaction decreases rapidly with distance because  the NaSrMn$_{2}$F$_{7}$ is an insulator. Therefore,  Heisenberg terms are calculated up to the third nearest neighbor($J_\mathrm{1}, J_\mathrm{2}$ and $J_\mathrm{3a}$). To this end, one has to regard the conventional unit-cell (88 atoms) with $4\times4\times4$ optimized Monkhorst-Pack k-point mesh. Then, $J_\mathrm{1}, J_\mathrm{2}$ and $J_\mathrm{3a}$ values can be obtained by calculating the total energy of different collinear spin configurations and mapping them to the Heisenberg model (Fig.~\ref{fig:Hzn} and Fig.~\ref{fig:J_fit}). There are two possible third nearest neighbor exchange interaction in this structure, $J_\text{3a}$ and $J_\text{3b}$\cite{KENNEDY1997, Wills2006, Sadeghi2015}.
Generally, in the cubic pyrochlore structures $J_\text{3b}\ll J_\text{3a}$. Therefore, $J_\text{3b}$ is approximated 0\cite{KENNEDY1997, Wills2006, Sadeghi2015}.

The nearest neighbor coupling is important for $B$ and anisotropic terms such as Dzyaloshinskii-Moriya coupling ($D$) and single-ion anisotropy strength ($\Delta$). Therefore, the primitive cell (22 atoms) with $6\times6\times6$
optimized Monkhorst-Pack k-point mesh must be regarded.
The strength of $B$ can be determined via total energy of noncollinear spin configurations in the absence of spin-orbit coupling (GGA+$U$).
In this case, if one starts from the all-in/all-out spin configuration (where all spins are oriented towards in/out center of tetrahedra) and if one rotates the spins so that $\vec{S_{1}}+\vec{S_{2}}+\vec{S_{3}}+\vec{S_{4}}=0$, it can be shown that Heisenberg term is degenerated~\cite{Sadeghi2015}.
The total energy difference of different noncollinear spin configurations can be attributed to $B$~\cite{Sadeghi2015}.

With the uniform rotation of magnetic moments in the absence of SOC, the isotropic terms of the spin Hamiltonian are degenerated. By considering the spin-orbit coupling (GGA+$U$+SOC), the total energy difference of the various noncollinear spin configurations is entirely attributed to $D$ provided that the SIA is negligible~\cite{Sadeghi2015}.
Given that Mn$^{+2}$ ends in 3d$^{5}$, it is expected that SIA has a very small contribution. To calculate this term, one has to examine two different noncollinear spin configurations. In the first spin configuration, one magnetic moment points outward from the center of the tetrahedra. The other three moments point inward toward the center of the tetrahedra. For the second spin configuration, the magnetic moments of the first configuration rotate about 120$^{\circ}$ around the $Z$ axis. It can be demonstrated that for these two spin configurations in the absence of SOC, isotropic terms do not change. In the presence of SOC, the $D$ remains unchanged and the total energy difference between these two spin configurations is entirely attributed to $\Delta$~\cite{Xiang2011}.
%%Table. 1%%%%%%%%%%%%%%%%%%%%%%%%%%%%%%%%%%%%%%%%%%%%%%%%%%%%%%%%%%%%%%%%%%%%%%%%%%%%%%%%%%%%%%%%%%%%%%%%%%%%%%%
\begin{table*} [!htp]
  \centering
  \caption{
  For various $U_\text{eff}$ values, Heisenberg couplings up to third neighbors have been determined. A negative value denotes anti-ferromagnetic exchange inetraction, whereas a positive value denotes ferromagnetic exchange interaction.
T$\text{f}$ can also be determined via MC simulations. To account for the effect of random A-site occupations, $\delta J$$=$0.13(meV) can be regarded as a weak bond disorder.}
      \begin{tabular}{ccccc}
    \hline
   $U_\text{eff}$ (eV) & $J_\text{1}$ (meV) & $J_\text{2}$ (meV) & $J_\text{3}$ (meV) &T$_\text{f}$ (K)\\
  \hline

4.58 & $-$1.97  & $-$0.03  & 0.02 & 1.8\\
4.00 & $-$2.33  & $-$0.04  & 0.03 &2.6\\
%3.30 & $-$3.46  & $-$0.01  & $-$0.02 & 0.005& $-$89.19\\
%3.58 & $-$3.22  & $-$0.01  & $-$0.02 & 0.002 &$-$79.90 \\
%4.58 & $-$1.97  & $-$0.03  & 0.02 &? \\
   \hline
   EXP & & & &   ~2.5~\cite{sanders2016nasrmn2f7}\\
     \hline
   \label{tab1}
 \end{tabular}
\end{table*}
%%%%%%%%%%%%%%%%%%%%%%%%%%%%%%%%%%%%%%%%%%%%%%%%%%%%%%%%%%%%%%%%%%%%

The derivation Heisenberg term and Other terms of Eq.~\eqref{H} have been done by the full-potential linearized augmented plane wave
(FPLAW) method as implemented in the FLEUR~\cite{fleur} code and Quantum Espresso (QE)~\cite{Giannozzi_2009, Giannozzi_2017} code which employs the plane wave basis set, respectively. For the
exchange-correlation energy, I employ the Perdew Burke Ernzerhof
parametrization of the generalized gradient approximation
(GGA)~\cite{Perdew1996}. The GGA+$U$ approximation~\cite{Cococcioni2005} is used to account
for the on-site Coulomb interaction for $3d$ orbitals of Mn atoms. The optimized cut-off of wave function expansion in the
interstitial region is set to
$k_{\mathrm{max}}=4.2\,\ \mathrm{a.u.}^{-1}$. The muffin-tin radius of
Na, Sr, Mn and F atoms are set to 2.6, 2.8, 2.3 and 1.4 a.u., respectively. 
To increase the accuracy of the calculations, I have considered the 3s and 3p orbitals of the Mn as semicore electrons (FLAPW+LO).
The effective $U$ parameter ($U_\text{eff}$=$U-J_\text{H}$), where $U$
and $J_\text{H}$ are screened on-site Coulomb and Hund exchange interactions,
is calculated by the Density Functional Perturbation Theory (DFPT)~\cite{Timrov2018}
using the QE code. The optimized 40 Ry and 400 Ry cutoff have been considered for expanding wavefunction and charge density in plane wave, respectively. The approximation electron-ion interaction has been controlled through GBRV ultra-soft pseudo-potential~\cite{Vanderbilt2014}.
In order to investigate the effect of random distribution on A-site,  one considers a conventional unitcell with 88 atoms. For this unitcell, 97 different configurations with different A-site distributions can be defined by using the supercell program~\cite{Okhotnikov2016}. Each of these distributions  has a specific and unique symmetry. For  calculations of the total energy for each configuration, QE code has been used in GGA approach.

To capture the critical
properties and low-temperature spacial spin  arrangement of NaSrMn$_{2}$F$_{7}$, one performs MC
simulation. The effect of charge disorder has been regarded in this simulation. Since SG materials have multiple local minimums in their energy landscape, reaching to a global minimum specially at low-temperature is not possible. To tackle this problem, the replica exchange method~\cite{Hukushima1996} should be employed as
implemented in the ESpinS package~\cite{REZAEI2022110947}. 
The Gaussian distribution has been considered as the  particular choice of disorder. 
In the replica exchange algorithm, the high-temperature configurations are swapped for the low-temperature configurations so that the system reaches thermal equilibrium. The condition for the system to reach equilibrium is that the MC steps should be sufficiently selected so that, first, the temperature behavior of the specific heat and energy of the system is smooth and, second, the position of the specific heat peak changes slightly by changing the seed and third, the acceptance ratio of MC steps at each temperature reaches at least to 65 percent.
The  linear size of the simulation cell (L$=$16) is used to ensure the presence of phase transition. In this way, there are  N$\times \mathrm{L}^{3}$ spins in the three-dimensional lattice where N is the number of spins
(N$=$4 for one tetrahedron).
%Fig.3 ##################################################################%%%%%%%%%%%%%%%%%%%%%%%%%%%%%%%%%%%%%%%%%%%%%%%%%%%%%%%%%%%
\begin{figure} [!htp]
    \centering
    \includegraphics[width=1.0\columnwidth]{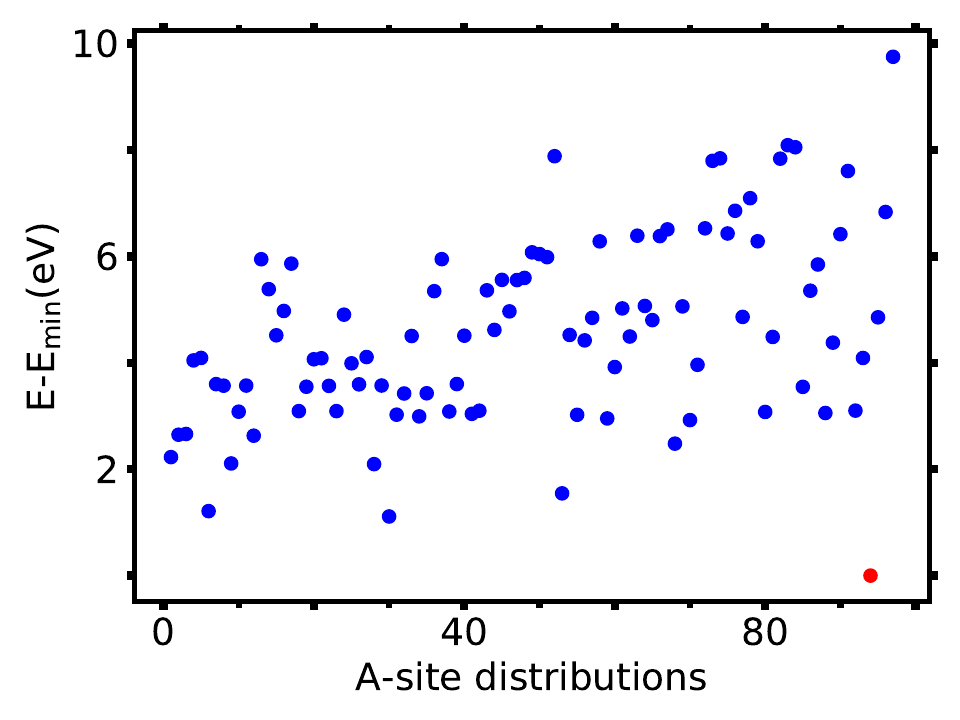}
    \caption{(Color online)   \textit{Ab initio} calculations of total energy for different A-site distributions in GGA approach. The red circle which has minimum total energy is related to homogeneous distribution in which Na and Sr occupy A-site tetrahedral with two by two ratio. The total energy of homogeneous distribution has been considered as E$_\text{min}$. It should be noted that primitive cell for calculating exchange paremeters beyond Heisenberg term can be made only for the homogeneous distribution.}
\label{fig:tfe}
\end{figure}
%%%%%%%%%%%%%%%%%%%%%%%%%%%%%%%%%%%%%%%%%%%%%%%%%%%%%%%%%%%%%%%%%%%%%%%%%%%%%%%%%%%%%%%%%%%%%%%%%%%%%%%%%%%%%%%%%%%%%%%%%%%%%%55
For MC calculations, when $\delta J$ is considered, the calculations are performed for ten different starting points (different seeds) and then the final result is expressed as the average of them.
3$\times10^{6}$ MC steps per
spin at each temperature are performed for the thermal equilibrium
and data collection, respectively. To reduce the correlation between
the successive data, measurements are done after skipping every 10 MC steps.
%Fig.4%%%%%%%%%%%%%%%%%%%%%%%%%%%%%%%%%%%%%%%%%%%%%%%%%%%%%%%%%%5%%%%%%%%%%%%%%%%%%%%%%%%%%%%%%%%%%%%%%%%%%%%%%
\begin{figure*}[!htp]
    \centering
    \includegraphics[width=0.9\columnwidth]{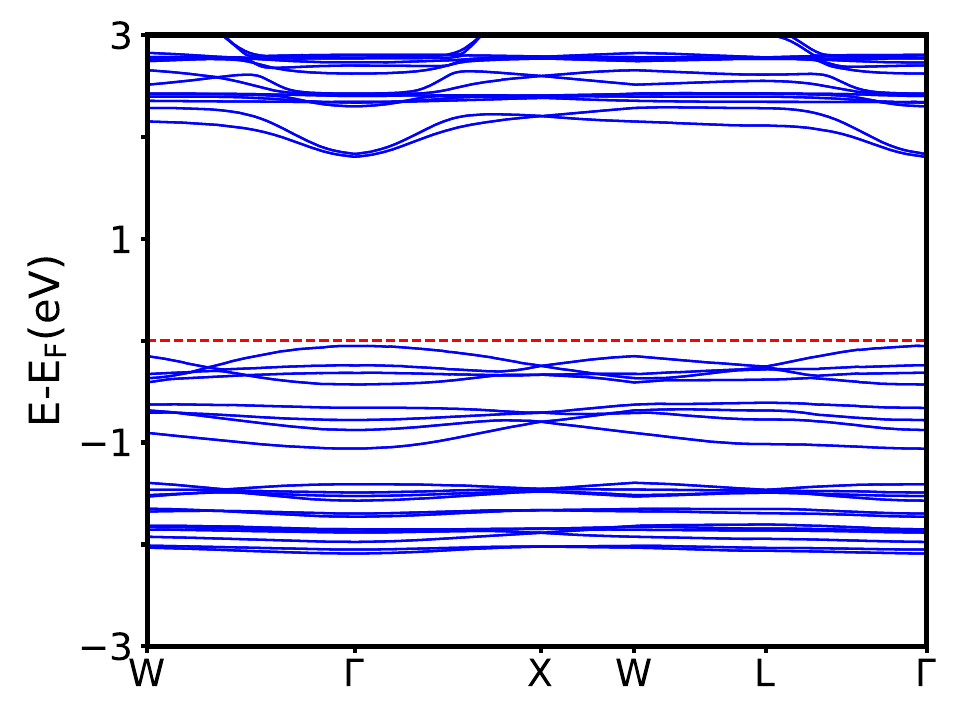}
    \hspace*{2cm}
    \includegraphics[width=0.9\columnwidth]{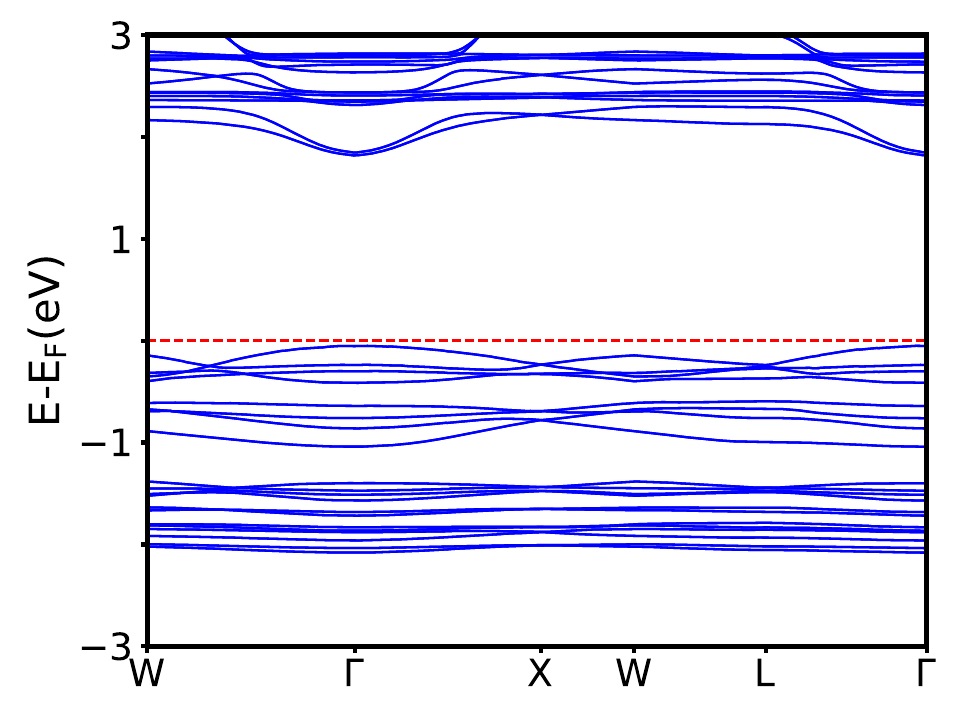}
    
   \caption{(Color online)Obtained the band structure of
NaSrMn$_{2}$F$_{7}$ in \textbf{(Left)}GGA and \textbf{(Right)}GGA+SOC approaches, respectively. The Fermi energy is set to zero. The all-in/all-out magnetic configuration (noncollinear spin texture) has been considered for these calculations. The figures exhibit that the SOC has an ignorable effect on energy bands near the Fermi level.}
   \label{fig:bands}
\end{figure*}
 %%%%%%%%%%%%%%%%%%%%%%%%%%%%%%%%%%%%%%%%%%%%%%%%%%%%%%%%%%%%%%%%%%%%%%%%%%%%%%%%%%%%%%%%%%%%%%%%%%%%%%%%
%%Fig. 5%%%%%%%%%%%%%%%%%%%%%%%%%%%%%%%%%%%%%%%%%%%%%%%%%%%%%%%%%%%%%%%%%%%%%%%%%%%%%%%%%%%%%%%%%%%%%%%%%%%%%%%%%%%%%%%%%
\begin{figure}[!htp]
    \centering
    \includegraphics[width=0.9\columnwidth]{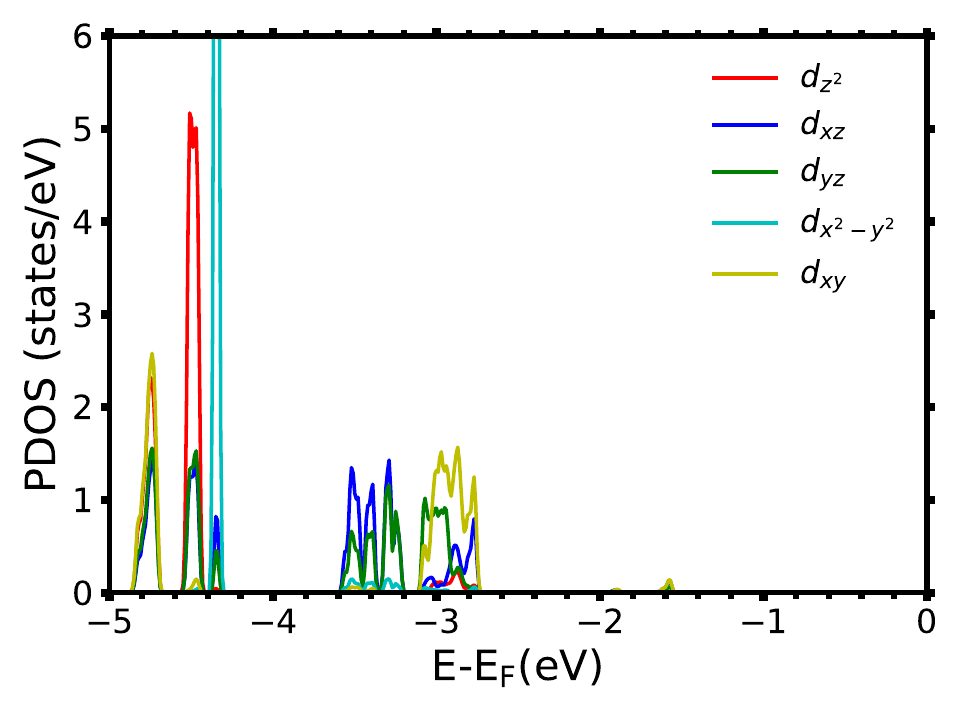}
    \caption{(Color online) PDOS calculated for AFM-ordered NaSrMn$_{2}$F$_{7}$ using the GGA functional. It illustrates the local 3$d$ DOS of a single Mn atom for majority spins. The obtained Lowdin Charge analysis indicates that the majority spin levels are fully occupied and minority spin levels are negligibly occupied. That's why the 3$d$ DOS for minority of spins has not been exhibited.
    It should be noted that this calculation has been done by using QE code~\cite{Giannozzi_2009, Giannozzi_2017}.}
    \label{fig:pdos_afm}
\end{figure}
 %%%%%%%%%%%%%%%%%%%%%%%%%%%%%%%%%%%%%%%%%%%%%%%%%%%%%%%%%%%%%%%%%%%%%%%%%%%%%%%%%%%%%%%%%%%%%%%%%%%%%%%%%%%%%%%%%%%%%%
%Results and Discussion%%%%%%%%%%%%%%%%%%%%%%%%%%%%%%%%%%%%%%%%%%%%%%%%%%%%%%%%%
\section{Results and Discussion}
\label{sec:results}
\subsection{Different distributions of Na-Sr on A-site}
A-site is completely randomly occupied by Na/Sr with equal concentrations. I take into account 97 various configurations with various A-site distributions with a distinct and special symmetry in order to examine the impact of varied A-site distributions on structural and magnetic properties.
First, I compute the total energy of all 97 distributions. The diagram reffig:tfe depicts how the different 97 configurations are arranged in terms of total energy.
In the GGA approach, I use the QE code for this calculation.
One distribution has the lowest total energy, according to Fig.~\ref{fig:tfe}. I call it homogeneous distribution because Na-Sr occupy A-site in such a way that there are two Sr and two Na in each A-site tetrahedral. For this distribution, one can only make primitive cells with 22 atoms.
Although the obtained band structure (Fig.~\ref{fig:bands}) suggests that the material is an insulator in the GGA approach, one needs to describe accurately the  electron-electron Coulomb correlation due to the 3d orbital of the Mn atom.
Therefore, one should apply the Hubbard potential to 3d electrons as an on-site Coulomb interaction.
Based on the \textit{ab initio} calculations (DFPT method), the effective Hubbard parameter $U_\text{eff}$ obtains 4.58~eV  for this system.
In this paper, I have taken into account $J_\text{H}$$\sim$1.0 eV as it is suggested for 3d materials~\cite{coey2013introduction, Vaugier2012}.
After determining the $U$ parameter, I optimize the primitive cell's lattice constants and ion locations. To eliminate bias, I examine FM order and maintain symmetry so that the final geometry has the same experimental symmetry and space group as the original geometry. The obtained results reveal that the F-Mn-F angle is distorted in the $<$111$>$ direction (see Fig.~\ref{fig:struct_pyr}). The estimated values are 95.48$^\circ$ and 84.51$^\circ$, which accord well with the experimental values (97.8$^\circ$ and 85.2$^\circ$). The optimized $x$ parameter and lattice constant are 0.3337 and 10.8930$\angstrom$, respectively, which are equivalent to the experimental values 0.3331 and 10.8012$\angstrom$.

%%%%%%%%%%%%%%%%%%%%%%%%%%%%%%%%%%%%%%%%%%%%%%%%%%%%%%%%%%%%%%%%%%%%%Magnetic properties%%%%%%%%%%%%%%%%%%%%%%%%%%%%%%%%%5
\subsection{Magnetic and electronic properties}
The quantitative derivation spin Hamiltonian (Eq.~\eqref{H}) is effective to explain or predict the magnetic properties of this pyrochlore. This information can answer the fundamental question about how a novel ground state emerges.
The results of calculations of different spin Hamiltonian terms for different Hubbard parameters are given in Table.~\ref{tab1}.
%%%%################################################Fig.6##################################################################
\begin{figure*} [!htp]
    \centering
\includegraphics[width=0.9\columnwidth]{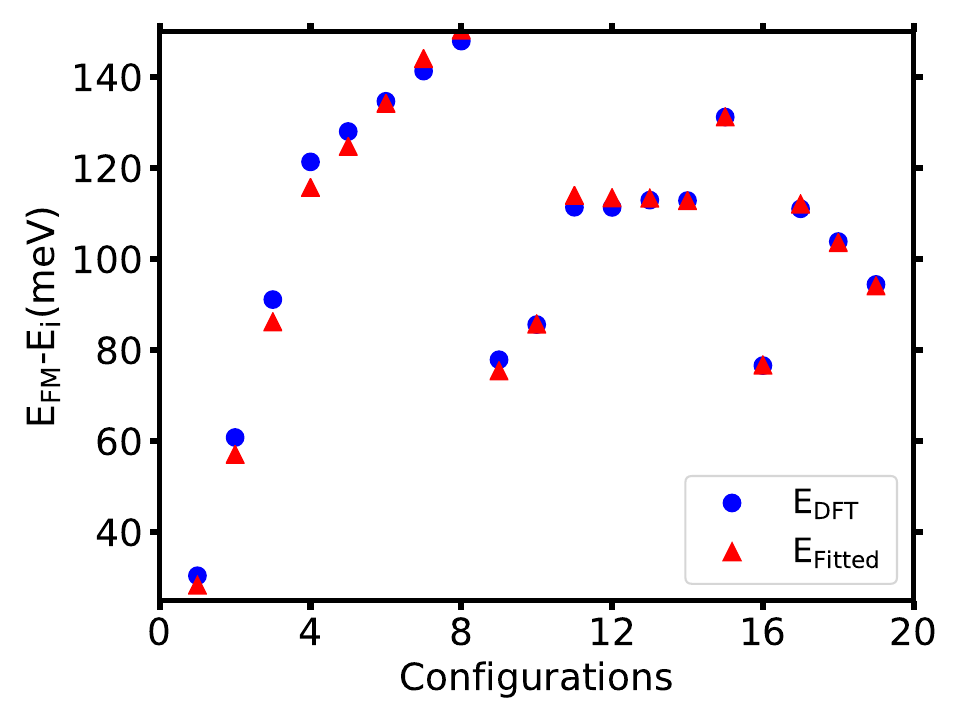}
    \hspace*{2cm}
    \includegraphics[width=0.9\columnwidth]{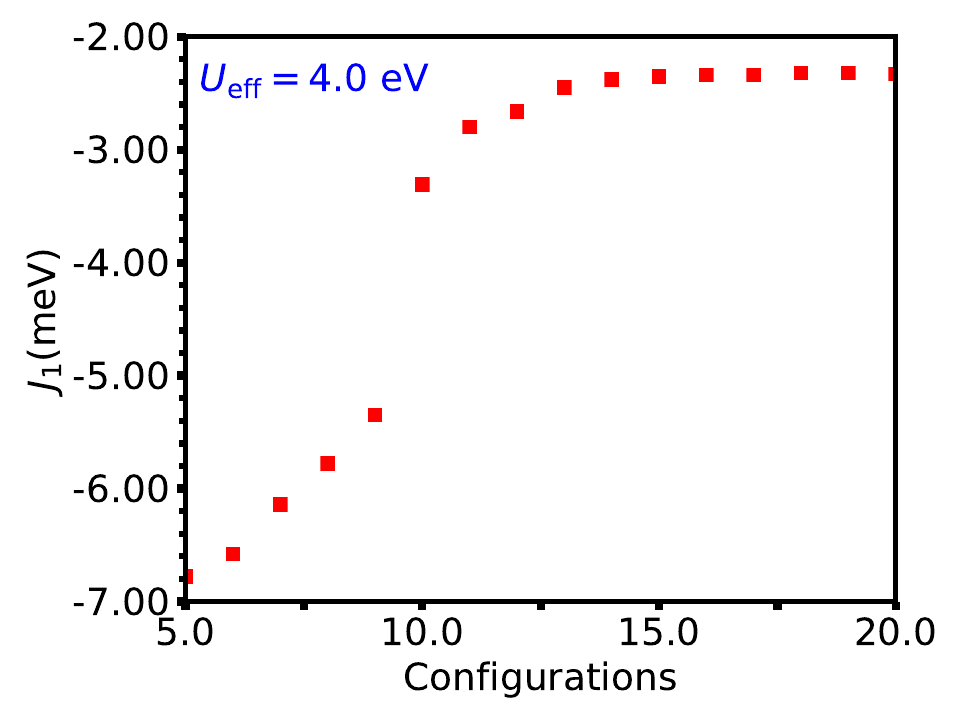}
    \caption{(Color online) $\bf(Left)$ Mapping total energy for various collinear spin configurations obtained by DFT+$U$ into the Heisenberg model. The error of fitting is 2\%.
    $\bf(right)$ Variation of $J_{1}$ with respect to different random collinear spin configurations. After 20 different configurations, $J_{1}$ trend has been constant. }
\label{fig:J_fit}
\end{figure*}
%%%%%%%%%%%%%%%%%%%%%%%%%%%%%%%%%%%%%%%%%%%%%%%%%%%%%%%%%%%%%%%%%%%%%%%%%%%%%%%%%%%%%%%%%%%%%%%%%%%%%
%Fig.7##################################################################%%%%%%%%%%%%%%%%%%%%%%%%%%%%%
\begin{figure*} [!htp]
    \centering
    \includegraphics[width=0.9\columnwidth]{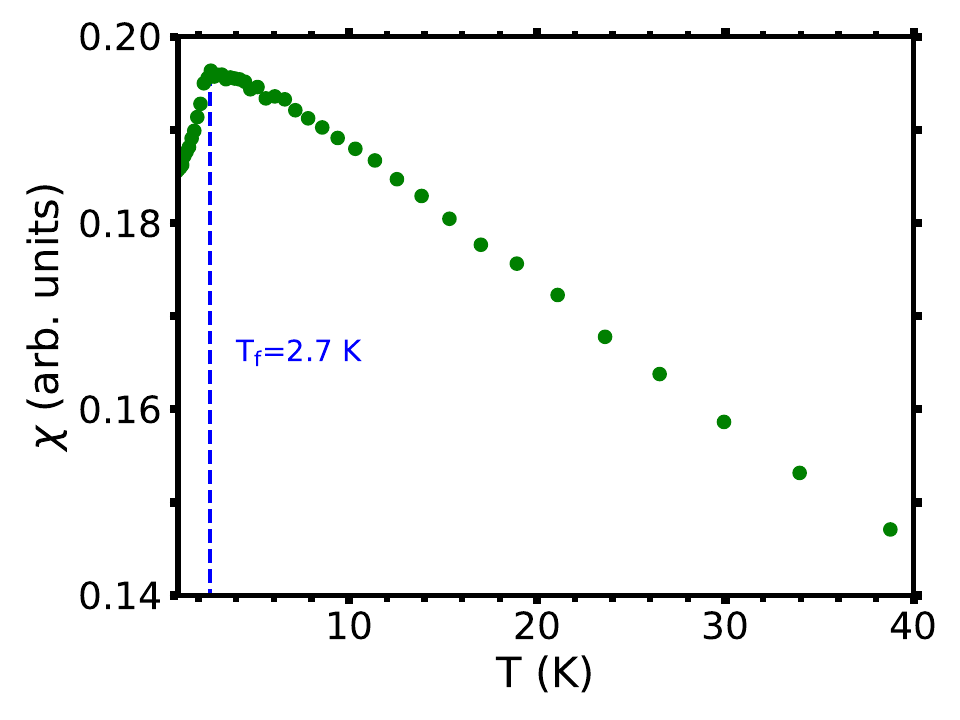}
    \hspace*{2cm}
    \includegraphics[width=0.9\columnwidth]{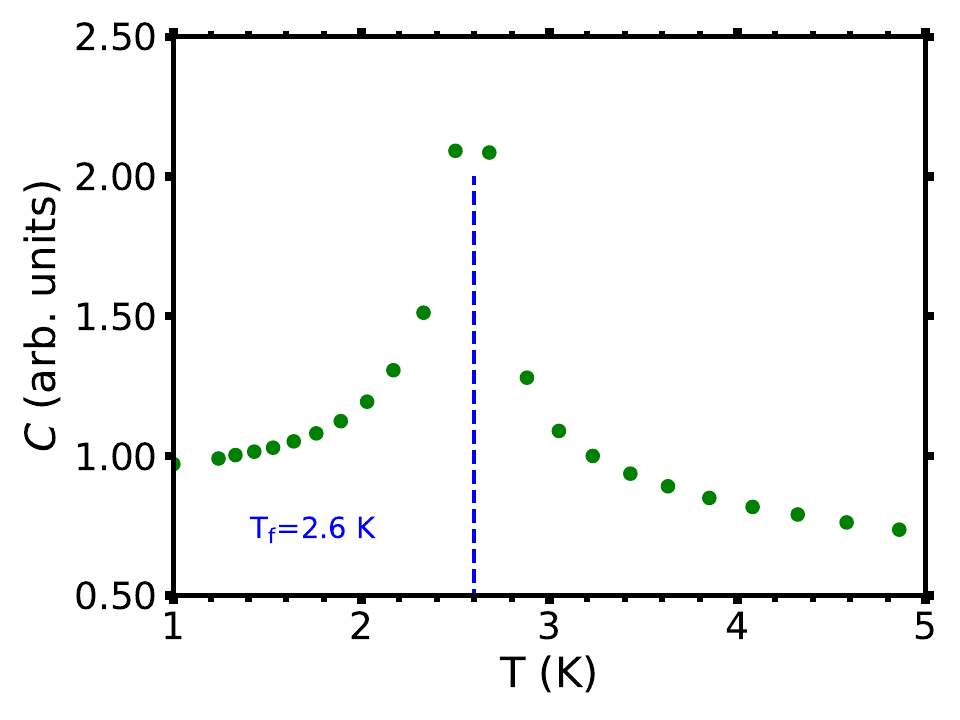}
    \caption{(Color online) The calculated $\bf(Left)$ magnetic specific heat and $\bf(right)$ magnetic susceptibility for $\delta J$$=$0.13 (meV). The obtained freezing temperature is similar to the experimental value. The temperature of the peak has been considered as T$_\text{f}$ because this point is where the break between high and low temperature data is defined.
    L=16 has been used as the size of the simulation lattice for these calculations.}
    \label{fig:termo-phase}
\end{figure*}
%%%%%%%%%%%%%%%%%%%%%%%%%%%%%%%%%%%%%%%%%%%%%%%%%%%%%%%%%%%%%%%%%%%%%%%%%%%%%%%%%%%%%%%%%%%%%%%%%%%%%%%
To ensure the exact value of $U_\text{eff}$, T$_\text{f}$ obtained from the MC simulation can be compared with the experimental value. 
As shown in Table.~\ref{tab1}, I have started from $U_\text{eff}$=4.58~eV and decreased it until the obtained T$_\text{f}$ is equal to the experimental value.
In this way, the optimum $U_\text{eff}$ is 4.0~eV. 

According to the chemical valence of different atoms in  NaSrMn$_{2}$F$_{7}$ compound, one expects that Na, Sr, Mn and F are in the +1, +2, +2 and -1 oxidation states, respectively. Therefore, the external shell of Mn$^{+2}$ is 3d$^{5}$ and
Hund rule predicts a spin S=5/2, and thereby a
magnetic moment 5 $\mu_{B}$ for it. The DFT calculations obtain
4.8 $\mu_{B}$, not far from the native ionic picture.
With this simple picture, one expects the quantum angular momentum of Mn to be zero and the magnetization of this compound originated only from spin part.
The DFT calculations give some residual magnetization 0.024 $\mu_{B}$, 0.0008 $\mu_{B}$ and  0.0005 $\mu_{B}$ for
F, Sr and Na atoms, respectively.  

To ensure that the bond disorder does not affect the crystal field and its consequences for higher order exchange interactions, other important and effective exchange couplings must be calculated.
DFT calculations show that other exchange parameters like $B$, $D$, and $\Delta$ are 3 to 4 times smaller than $J_{1}$. As a result, these interactions are negligible and have no bearing on the magnetic properties of the ground state. The orbital moment of the magnetic ion strongly influences the antisymmetric exchange interactions (DMI and SIA). The Mn orbital moment plays a very small role in the effective orbital of this material, according to experimental measurements~\cite{sanders2016nasrmn2f7}.
The orbital moment of each Mn ion obtained from \textit{ab initio} calculations (GGA+$U$+SOC) is 0.0001 $\mu_{B}$, which is a very small amount. It is possible to conclude that antisymmetric exchange interactions in this material are unimportant. Fig.~\ref{fig:bands} also compares band structures for noncollinear (all-in/all-out) magnetic configurations with and without SOC. As can be seen, the SOC has very little effect on the energy bands, especially near the Fermi level. As a result, the SOC effect in this material is very weak. The spin-projected DOS (PDOS) and Lowdin charge analysis calculations for a Mn atom confirm that the SOC effect is negligible (Fig.~\ref{fig:pdos_afm}). 
According to the Lowdin charge analysis, the total charge for majority and minority spin levels are $d_{z^{2}}$=0.9973, $d_{xz}$=0.9905, $d_{yz}$=0.9909, $d_{x^{2}-y^{2}}$=0.9973, $d_{xy}$=0.9927 and $d_{z^{2}}$=0.0348, $d_{xz}$=0.0723, $d_{yz}$=0.0707, $d_{x^{2}-y^{2}}$=0.0358, $d_{xy}$=0.0723, respectively. 
As a result, the majority spin levels are completely occupied, while the minority spin levels are barely occupied. It is reasonable to conclude that the orbital moment is quenched in this material. 
Fig.~\ref{fig:J_fit} shows how $J_{1}$  has been calculated. As one can see, the choosing number of different collinear magnetic configurations continues until $J_{1}$ reaches a constant value which is $-$2.33 (meV).
Also, Fig.~\ref{fig:J_fit} illustrates how the total energies from DFT+$U$ calculations are mapped to the Heisenberg model via the least-square fitting method. The error of fitting obtains 2\%.
The exchange parameters of the second ($J_{2}$) and third ($J_{3}$) nearest neighbors are very small.
Thus, A-site random occupation only changes $J_{1}$. One can estimate this weak bond disorder by using $\delta J=\sqrt \frac{3}{8}k_\text{B}$T$_\text{f}$~\cite{saunders2007spin}. Since the experimental value of the freezing temperature is 2.5 (K)~\cite{sanders2016nasrmn2f7}, $\delta J$ is equal to 0.13 (meV).
%%%%%%%%%%%%%%%%%%%%%%%%%%%%%%%%%%%%%%%%%%Critical properties%%%%%%%%%%%%%%%%%%%%%%%%%%%%%%%%%%%%%%%%%%%%%%%%%%%%%55
\subsection{Critical properties}
In the following, I perform MC simulations to find the optimal $U_\text{eff}$ and the \textit{ab initio} freezing temperature. The calculations are done taking into consideration $\delta J$ to account for the impact of A-site random occupancy on MC simulation.
The temperature behavior of the sample's specific heat ($C$) and magnetic susceptibility ($chi$) are shown in Fig.~\ref{fig:termo-phase}.
I consider the position of peak for $C$ and $\chi$ to define freezing temperature. For $\delta J$=0.13~(meV), T$_\text{f}$ obtains 2.6$\pm$0.1~K.
By decreasing the amount of $\delta J$, the freezing temperature decreases so that  for $\delta J$=0, no phase transition is observed up to 0.5~K.
Therefore, one can conclude that the $\delta J$ acts as a small perturbation and must account for the system SG phase transition at low-temperature by eliminating the degeneracy of the ground state~\cite{saunders2007spin}.
%%%%%%%%%%%%%%%%%%%%%%%%%%%%%%%%%%%%%Conclusion%%%%%%%%%%%%%%%%%%%%%%%%%%%%%%%%%%%%%%%%%%%%%%%%%%%%%%%%%%%%%%%%%%%%5
\section{Conclusion}
\label{sec:conclusion}
In conclusion, spin Hamiltonian for spin-glass pyrochlore NaSrMn$_{2}$F$_{7}$ were calculated using DFT+$U$+SOC. The spin Hamilton comprised isotropic and anisotropic concepts such nearest neighbor bi-quadratic, Dzyaloshinskii-Moriya, and single-ion anisotropy, as well as Heisenberg coupling up to third neighbors. 
Since the Hubbard parameter played a decisive role in describing the electron-electron correlation and consequently the magnetic properties, this value was determined via the DFPT method. Then, I reduced it to such an extent that the T$_\mathrm{f}$ obtained from the MC simulations equaled to the experimental T$_\mathrm{f}$.
The DFT computations revealed that the sole thing that matters most in determining this material's magnetic characteristics is $J_{1}$ (as nearest neighbor Heisenberg exchange). 
Following that, the specific heat and magnetic susceptibility temperature behavior were calculated using the obtained spin Hamiltonian from MC simulations.
Because Na-Sr random occupancy on A-site causes weak bond disorder($\delta J$), $\delta J$$=$0.13 (meV) was assumed to produce freezing temperature equals the experimental freezing temperature.
$\delta J$ behaves like a perturbation and is what causes the spin-glass phase transition at low temperatures because no phase transition can be seen even at low temperatures when $\delta J$$=$0.
To better comprehend the nature of fluctuations at low temperatures and their impact on electron-electron correlation, experimental data like inelastic neutron scattering may be useful.
%%%%%##################################################################################################################
\begin{acknowledgments}
J\"ulich Supercomputing Centre (JSC) is gratefully acknowledged by author as the
provider of needed computing facilities. 
The guidelines provided by Peter Kratzer, Marjana Le\v{z}ai\'{c} and Mojaba Alaei are highly valued and appreciated.
\end{acknowledgments}
%%%%%##################################################################################################################
%%%%%##################################################################################################################
%%%%%##################################################################################################################
%
%%%%%%%%%%%%%%%%%%%%%%%%%%%%%%%%%%%%%%%%%%%%%%%%%%%%%%%%%%%%%%%%%%%%%%%%%%%%%%%%%%%%%%%%%%%%%%%%%%%%%%%%%%%%%%%%%%%%%%%%%%%%%%%%%

\begin{thebibliography}{47}%
\makeatletter
\providecommand \@ifxundefined [1]{%
 \@ifx{#1\undefined}
}%
\providecommand \@ifnum [1]{%
 \ifnum #1\expandafter \@firstoftwo
 \else \expandafter \@secondoftwo
 \fi
}%
\providecommand \@ifx [1]{%
 \ifx #1\expandafter \@firstoftwo
 \else \expandafter \@secondoftwo
 \fi
}%
\providecommand \natexlab [1]{#1}%
\providecommand \enquote  [1]{``#1''}%
\providecommand \bibnamefont  [1]{#1}%
\providecommand \bibfnamefont [1]{#1}%
\providecommand \citenamefont [1]{#1}%
\providecommand \href@noop [0]{\@secondoftwo}%
\providecommand \href [0]{\begingroup \@sanitize@url \@href}%
\providecommand \@href[1]{\@@startlink{#1}\@@href}%
\providecommand \@@href[1]{\endgroup#1\@@endlink}%
\providecommand \@sanitize@url [0]{\catcode `\\12\catcode `\$12\catcode
  `\&12\catcode `\#12\catcode `\^12\catcode `\_12\catcode `\%12\relax}%
\providecommand \@@startlink[1]{}%
\providecommand \@@endlink[0]{}%
\providecommand \url  [0]{\begingroup\@sanitize@url \@url }%
\providecommand \@url [1]{\endgroup\@href {#1}{\urlprefix }}%
\providecommand \urlprefix  [0]{URL }%
\providecommand \Eprint [0]{\href }%
\providecommand \doibase [0]{http://dx.doi.org/}%
\providecommand \selectlanguage [0]{\@gobble}%
\providecommand \bibinfo  [0]{\@secondoftwo}%
\providecommand \bibfield  [0]{\@secondoftwo}%
\providecommand \translation [1]{[#1]}%
\providecommand \BibitemOpen [0]{}%
\providecommand \bibitemStop [0]{}%
\providecommand \bibitemNoStop [0]{.\EOS\space}%
\providecommand \EOS [0]{\spacefactor3000\relax}%
\providecommand \BibitemShut  [1]{\csname bibitem#1\endcsname}%
\let\auto@bib@innerbib\@empty
%</preamble>
\bibitem [{\citenamefont {Mydosh}(1993)}]{mydosh1993spin}%
  \BibitemOpen
  \bibfield  {author} {\bibinfo {author} {\bibfnamefont {J.~A.}\ \bibnamefont
  {Mydosh}},\ }\href@noop {} {\emph {\bibinfo {title} {Spin glasses: an
  experimental introduction}}}\ (\bibinfo  {publisher} {CRC Press},\ \bibinfo
  {year} {1993})\BibitemShut {NoStop}%
\bibitem [{\citenamefont {Binder}\ and\ \citenamefont
  {Young}(1986)}]{Binder1986}%
  \BibitemOpen
  \bibfield  {author} {\bibinfo {author} {\bibfnamefont {K.}~\bibnamefont
  {Binder}}\ and\ \bibinfo {author} {\bibfnamefont {A.~P.}\ \bibnamefont
  {Young}},\ }\href {\doibase 10.1103/RevModPhys.58.801} {\bibfield  {journal}
  {\bibinfo  {journal} {Rev. Mod. Phys.}\ }\textbf {\bibinfo {volume} {58}},\
  \bibinfo {pages} {801} (\bibinfo {year} {1986})}\BibitemShut {NoStop}%
\bibitem [{\citenamefont {Taniguchi}\ \emph {et~al.}(2009)\citenamefont
  {Taniguchi}, \citenamefont {Munenaka},\ and\ \citenamefont
  {Sato}}]{taniguchi2009spin}%
  \BibitemOpen
  \bibfield  {author} {\bibinfo {author} {\bibfnamefont {T.}~\bibnamefont
  {Taniguchi}}, \bibinfo {author} {\bibfnamefont {T.}~\bibnamefont {Munenaka}},
  \ and\ \bibinfo {author} {\bibfnamefont {H.}~\bibnamefont {Sato}},\ }in\
  \href {\doibase 10.1088/1742-6596/145/1/012017} {\emph {\bibinfo {booktitle}
  {Journal of Physics: Conference Series}}},\ Vol.\ \bibinfo {volume} {145}\
  (\bibinfo {organization} {IOP Publishing},\ \bibinfo {year} {2009})\ p.\
  \bibinfo {pages} {012017}\BibitemShut {NoStop}%
\bibitem [{\citenamefont {Shintani}\ and\ \citenamefont
  {Tanaka}(2006)}]{shintani2006frustration}%
  \BibitemOpen
  \bibfield  {author} {\bibinfo {author} {\bibfnamefont {H.}~\bibnamefont
  {Shintani}}\ and\ \bibinfo {author} {\bibfnamefont {H.}~\bibnamefont
  {Tanaka}},\ }\href {\doibase 10.1038/nphys235} {\bibfield  {journal}
  {\bibinfo  {journal} {Nature Physics}\ }\textbf {\bibinfo {volume} {2}},\
  \bibinfo {pages} {200} (\bibinfo {year} {2006})}\BibitemShut {NoStop}%
\bibitem [{\citenamefont {Sanders}\ \emph {et~al.}(2016)\citenamefont
  {Sanders}, \citenamefont {Krizan}, \citenamefont {Plumb}, \citenamefont
  {McQueen},\ and\ \citenamefont {Cava}}]{sanders2016nasrmn2f7}%
  \BibitemOpen
  \bibfield  {author} {\bibinfo {author} {\bibfnamefont {M.}~\bibnamefont
  {Sanders}}, \bibinfo {author} {\bibfnamefont {J.}~\bibnamefont {Krizan}},
  \bibinfo {author} {\bibfnamefont {K.}~\bibnamefont {Plumb}}, \bibinfo
  {author} {\bibfnamefont {T.}~\bibnamefont {McQueen}}, \ and\ \bibinfo
  {author} {\bibfnamefont {R.}~\bibnamefont {Cava}},\ }\href {\doibase
  10.1088/1361-648x/29/4/045801} {\bibfield  {journal} {\bibinfo  {journal}
  {Journal of Physics: Condensed Matter}\ }\textbf {\bibinfo {volume} {29}},\
  \bibinfo {pages} {045801} (\bibinfo {year} {2016})}\BibitemShut {NoStop}%
\bibitem [{\citenamefont {Kawamura}\ and\ \citenamefont
  {Taniguchi}(2015)}]{KAWAMURA20151}%
  \BibitemOpen
  \bibfield  {author} {\bibinfo {author} {\bibfnamefont {H.}~\bibnamefont
  {Kawamura}}\ and\ \bibinfo {author} {\bibfnamefont {T.}~\bibnamefont
  {Taniguchi}},\ }\href {\doibase https://doi.org/10.1016/bs.hmm.2015.08.001}
  {\emph {\bibinfo {title} {Chapter 1 - Spin Glasses}}},\ edited by\ \bibinfo
  {editor} {\bibfnamefont {K.}~\bibnamefont {Buschow}},\ \bibinfo {series}
  {Handbook of Magnetic Materials}, Vol.~\bibinfo {volume} {24}\ (\bibinfo
  {publisher} {Elsevier},\ \bibinfo {year} {2015})\ pp.\ \bibinfo {pages} {1 --
  137}\BibitemShut {NoStop}%
\bibitem [{\citenamefont {Vincent}\ and\ \citenamefont
  {Dupuis}(2018)}]{vincent2018spin}%
  \BibitemOpen
  \bibfield  {author} {\bibinfo {author} {\bibfnamefont {E.}~\bibnamefont
  {Vincent}}\ and\ \bibinfo {author} {\bibfnamefont {V.}~\bibnamefont
  {Dupuis}},\ }in\ \href@noop {} {\emph {\bibinfo {booktitle} {Frustrated
  Materials and Ferroic Glasses}}}\ (\bibinfo  {publisher} {Springer},\
  \bibinfo {year} {2018})\ pp.\ \bibinfo {pages} {31--56}\BibitemShut {NoStop}%
\bibitem [{\citenamefont {Mitsumoto}\ \emph {et~al.}(2020)\citenamefont
  {Mitsumoto}, \citenamefont {Hotta},\ and\ \citenamefont
  {Yoshino}}]{Mitsumoto2020}%
  \BibitemOpen
  \bibfield  {author} {\bibinfo {author} {\bibfnamefont {K.}~\bibnamefont
  {Mitsumoto}}, \bibinfo {author} {\bibfnamefont {C.}~\bibnamefont {Hotta}}, \
  and\ \bibinfo {author} {\bibfnamefont {H.}~\bibnamefont {Yoshino}},\ }\href
  {\doibase 10.1103/PhysRevLett.124.087201} {\bibfield  {journal} {\bibinfo
  {journal} {Phys. Rev. Lett.}\ }\textbf {\bibinfo {volume} {124}},\ \bibinfo
  {pages} {087201} (\bibinfo {year} {2020})}\BibitemShut {NoStop}%
\bibitem [{\citenamefont {Shinaoka}\ \emph {et~al.}(2011)\citenamefont
  {Shinaoka}, \citenamefont {Tomita},\ and\ \citenamefont
  {Motome}}]{Shinaoka2011}%
  \BibitemOpen
  \bibfield  {author} {\bibinfo {author} {\bibfnamefont {H.}~\bibnamefont
  {Shinaoka}}, \bibinfo {author} {\bibfnamefont {Y.}~\bibnamefont {Tomita}}, \
  and\ \bibinfo {author} {\bibfnamefont {Y.}~\bibnamefont {Motome}},\ }\href
  {\doibase 10.1103/PhysRevLett.107.047204} {\bibfield  {journal} {\bibinfo
  {journal} {Phys. Rev. Lett.}\ }\textbf {\bibinfo {volume} {107}},\ \bibinfo
  {pages} {047204} (\bibinfo {year} {2011})}\BibitemShut {NoStop}%
\bibitem [{\citenamefont {Subramanian}\ \emph {et~al.}(1983)\citenamefont
  {Subramanian}, \citenamefont {Aravamudan},\ and\ \citenamefont {{Subba
  Rao}}}]{SUBRAMANIAN198355}%
  \BibitemOpen
  \bibfield  {author} {\bibinfo {author} {\bibfnamefont {M.}~\bibnamefont
  {Subramanian}}, \bibinfo {author} {\bibfnamefont {G.}~\bibnamefont
  {Aravamudan}}, \ and\ \bibinfo {author} {\bibfnamefont {G.}~\bibnamefont
  {{Subba Rao}}},\ }\href {\doibase
  https://doi.org/10.1016/0079-6786(83)90001-8} {\bibfield  {journal} {\bibinfo
   {journal} {Progress in Solid State Chemistry}\ }\textbf {\bibinfo {volume}
  {15}},\ \bibinfo {pages} {55 } (\bibinfo {year} {1983})}\BibitemShut
  {NoStop}%
\bibitem [{\citenamefont {Gardner}\ \emph {et~al.}(2010)\citenamefont
  {Gardner}, \citenamefont {Gingras},\ and\ \citenamefont
  {Greedan}}]{Gardner2010}%
  \BibitemOpen
  \bibfield  {author} {\bibinfo {author} {\bibfnamefont {J.~S.}\ \bibnamefont
  {Gardner}}, \bibinfo {author} {\bibfnamefont {M.~J.~P.}\ \bibnamefont
  {Gingras}}, \ and\ \bibinfo {author} {\bibfnamefont {J.~E.}\ \bibnamefont
  {Greedan}},\ }\href {\doibase 10.1103/RevModPhys.82.53} {\bibfield  {journal}
  {\bibinfo  {journal} {Rev. Mod. Phys.}\ }\textbf {\bibinfo {volume} {82}},\
  \bibinfo {pages} {53} (\bibinfo {year} {2010})}\BibitemShut {NoStop}%
\bibitem [{\citenamefont {Lacroix}\ \emph {et~al.}(2011)\citenamefont
  {Lacroix}, \citenamefont {Mendels},\ and\ \citenamefont
  {Mila}}]{lacroix2011introduction}%
  \BibitemOpen
  \bibfield  {author} {\bibinfo {author} {\bibfnamefont {C.}~\bibnamefont
  {Lacroix}}, \bibinfo {author} {\bibfnamefont {P.}~\bibnamefont {Mendels}}, \
  and\ \bibinfo {author} {\bibfnamefont {F.}~\bibnamefont {Mila}},\ }\href@noop
  {} {\emph {\bibinfo {title} {Introduction to frustrated magnetism: materials,
  experiments, theory}}},\ Vol.\ \bibinfo {volume} {164}\ (\bibinfo
  {publisher} {Springer Science \& Business Media},\ \bibinfo {year}
  {2011})\BibitemShut {NoStop}%
\bibitem [{\citenamefont {Reig-i Plessis}\ and\ \citenamefont
  {Hallas}(2021)}]{Reig-i-Plessis2021}%
  \BibitemOpen
  \bibfield  {author} {\bibinfo {author} {\bibfnamefont {D.}~\bibnamefont
  {Reig-i Plessis}}\ and\ \bibinfo {author} {\bibfnamefont {A.~M.}\
  \bibnamefont {Hallas}},\ }\href {\doibase 10.1103/PhysRevMaterials.5.030301}
  {\bibfield  {journal} {\bibinfo  {journal} {Phys. Rev. Materials}\ }\textbf
  {\bibinfo {volume} {5}},\ \bibinfo {pages} {030301} (\bibinfo {year}
  {2021})}\BibitemShut {NoStop}%
\bibitem [{\citenamefont {Harris}\ \emph {et~al.}(1997)\citenamefont {Harris},
  \citenamefont {Bramwell}, \citenamefont {McMorrow}, \citenamefont {Zeiske},\
  and\ \citenamefont {Godfrey}}]{Harris1997}%
  \BibitemOpen
  \bibfield  {author} {\bibinfo {author} {\bibfnamefont {M.~J.}\ \bibnamefont
  {Harris}}, \bibinfo {author} {\bibfnamefont {S.~T.}\ \bibnamefont
  {Bramwell}}, \bibinfo {author} {\bibfnamefont {D.~F.}\ \bibnamefont
  {McMorrow}}, \bibinfo {author} {\bibfnamefont {T.}~\bibnamefont {Zeiske}}, \
  and\ \bibinfo {author} {\bibfnamefont {K.~W.}\ \bibnamefont {Godfrey}},\
  }\href {\doibase 10.1103/PhysRevLett.79.2554} {\bibfield  {journal} {\bibinfo
   {journal} {Phys. Rev. Lett.}\ }\textbf {\bibinfo {volume} {79}},\ \bibinfo
  {pages} {2554} (\bibinfo {year} {1997})}\BibitemShut {NoStop}%
\bibitem [{\citenamefont {Slobinsky}\ \emph {et~al.}(2021)\citenamefont
  {Slobinsky}, \citenamefont {Pili}, \citenamefont {Baglietto}, \citenamefont
  {Grigera},\ and\ \citenamefont {Borzi}}]{slobinsky2021monopole}%
  \BibitemOpen
  \bibfield  {author} {\bibinfo {author} {\bibfnamefont {D.}~\bibnamefont
  {Slobinsky}}, \bibinfo {author} {\bibfnamefont {L.}~\bibnamefont {Pili}},
  \bibinfo {author} {\bibfnamefont {G.}~\bibnamefont {Baglietto}}, \bibinfo
  {author} {\bibfnamefont {S.}~\bibnamefont {Grigera}}, \ and\ \bibinfo
  {author} {\bibfnamefont {R.}~\bibnamefont {Borzi}},\ }\href {\doibase
  https://doi.org/10.1038/s42005-021-00552-0} {\bibfield  {journal} {\bibinfo
  {journal} {Communications Physics}\ }\textbf {\bibinfo {volume} {4}},\
  \bibinfo {pages} {1} (\bibinfo {year} {2021})}\BibitemShut {NoStop}%
\bibitem [{\citenamefont {Moessner}\ and\ \citenamefont
  {Sondhi}(2003)}]{moessner2003theory}%
  \BibitemOpen
  \bibfield  {author} {\bibinfo {author} {\bibfnamefont {R.}~\bibnamefont
  {Moessner}}\ and\ \bibinfo {author} {\bibfnamefont {S.~L.}\ \bibnamefont
  {Sondhi}},\ }\href {\doibase 10.1103/PhysRevB.68.064411} {\bibfield
  {journal} {\bibinfo  {journal} {Physical Review B}\ }\textbf {\bibinfo
  {volume} {68}},\ \bibinfo {pages} {064411} (\bibinfo {year}
  {2003})}\BibitemShut {NoStop}%
\bibitem [{\citenamefont {Marlton}\ \emph {et~al.}(2021)\citenamefont
  {Marlton}, \citenamefont {Zhang}, \citenamefont {Zhang}, \citenamefont
  {Proffen}, \citenamefont {Ling},\ and\ \citenamefont
  {Kennedy}}]{marlton2021lattice}%
  \BibitemOpen
  \bibfield  {author} {\bibinfo {author} {\bibfnamefont {F.~P.}\ \bibnamefont
  {Marlton}}, \bibinfo {author} {\bibfnamefont {Z.}~\bibnamefont {Zhang}},
  \bibinfo {author} {\bibfnamefont {Y.}~\bibnamefont {Zhang}}, \bibinfo
  {author} {\bibfnamefont {T.~E.}\ \bibnamefont {Proffen}}, \bibinfo {author}
  {\bibfnamefont {C.~D.}\ \bibnamefont {Ling}}, \ and\ \bibinfo {author}
  {\bibfnamefont {B.~J.}\ \bibnamefont {Kennedy}},\ }\href {\doibase
  10.1021/acs.chemmater.0c04515} {\bibfield  {journal} {\bibinfo  {journal}
  {Chemistry of Materials}\ } (\bibinfo {year} {2021}),\
  10.1021/acs.chemmater.0c04515}\BibitemShut {NoStop}%
\bibitem [{\citenamefont {Yahne}\ \emph {et~al.}(2021)\citenamefont {Yahne},
  \citenamefont {Pereira}, \citenamefont {Jaubert}, \citenamefont {Sanjeewa},
  \citenamefont {Powell}, \citenamefont {Kolis}, \citenamefont {Xu},
  \citenamefont {Enjalran}, \citenamefont {Gingras},\ and\ \citenamefont
  {Ross}}]{yahne2021understanding}%
  \BibitemOpen
  \bibfield  {author} {\bibinfo {author} {\bibfnamefont {D.}~\bibnamefont
  {Yahne}}, \bibinfo {author} {\bibfnamefont {D.}~\bibnamefont {Pereira}},
  \bibinfo {author} {\bibfnamefont {L.}~\bibnamefont {Jaubert}}, \bibinfo
  {author} {\bibfnamefont {L.}~\bibnamefont {Sanjeewa}}, \bibinfo {author}
  {\bibfnamefont {M.}~\bibnamefont {Powell}}, \bibinfo {author} {\bibfnamefont
  {J.}~\bibnamefont {Kolis}}, \bibinfo {author} {\bibfnamefont
  {G.}~\bibnamefont {Xu}}, \bibinfo {author} {\bibfnamefont {M.}~\bibnamefont
  {Enjalran}}, \bibinfo {author} {\bibfnamefont {M.}~\bibnamefont {Gingras}}, \
  and\ \bibinfo {author} {\bibfnamefont {K.}~\bibnamefont {Ross}},\ }\href@noop
  {} {\bibfield  {journal} {\bibinfo  {journal} {arXiv preprint
  arXiv:2101.08361}\ } (\bibinfo {year} {2021})}\BibitemShut {NoStop}%
\bibitem [{\citenamefont {Shi}\ \emph {et~al.}(2021)\citenamefont {Shi},
  \citenamefont {Nisoli},\ and\ \citenamefont {Chern}}]{shi2021ice}%
  \BibitemOpen
  \bibfield  {author} {\bibinfo {author} {\bibfnamefont {Y.}~\bibnamefont
  {Shi}}, \bibinfo {author} {\bibfnamefont {C.}~\bibnamefont {Nisoli}}, \ and\
  \bibinfo {author} {\bibfnamefont {G.-W.}\ \bibnamefont {Chern}},\ }\href
  {\doibase 10.1063/5.0046083} {\bibfield  {journal} {\bibinfo  {journal}
  {Applied Physics Letters}\ }\textbf {\bibinfo {volume} {118}},\ \bibinfo
  {pages} {122407} (\bibinfo {year} {2021})}\BibitemShut {NoStop}%
\bibitem [{\citenamefont {Fennell}\ \emph {et~al.}(2009)\citenamefont
  {Fennell}, \citenamefont {Deen}, \citenamefont {Wildes}, \citenamefont
  {Schmalzl}, \citenamefont {Prabhakaran}, \citenamefont {Boothroyd},
  \citenamefont {Aldus}, \citenamefont {McMorrow},\ and\ \citenamefont
  {Bramwell}}]{fennell2009magnetic}%
  \BibitemOpen
  \bibfield  {author} {\bibinfo {author} {\bibfnamefont {T.}~\bibnamefont
  {Fennell}}, \bibinfo {author} {\bibfnamefont {P.}~\bibnamefont {Deen}},
  \bibinfo {author} {\bibfnamefont {A.}~\bibnamefont {Wildes}}, \bibinfo
  {author} {\bibfnamefont {K.}~\bibnamefont {Schmalzl}}, \bibinfo {author}
  {\bibfnamefont {D.}~\bibnamefont {Prabhakaran}}, \bibinfo {author}
  {\bibfnamefont {A.}~\bibnamefont {Boothroyd}}, \bibinfo {author}
  {\bibfnamefont {R.}~\bibnamefont {Aldus}}, \bibinfo {author} {\bibfnamefont
  {D.}~\bibnamefont {McMorrow}}, \ and\ \bibinfo {author} {\bibfnamefont
  {S.}~\bibnamefont {Bramwell}},\ }\href {\doibase
  https://doi.org/10.1126/science.1177582} {\bibfield  {journal} {\bibinfo
  {journal} {Science}\ }\textbf {\bibinfo {volume} {326}},\ \bibinfo {pages}
  {415} (\bibinfo {year} {2009})}\BibitemShut {NoStop}%
\bibitem [{\citenamefont {Bramwell}\ and\ \citenamefont
  {Gingras}(2001)}]{bramwell2001spin}%
  \BibitemOpen
  \bibfield  {author} {\bibinfo {author} {\bibfnamefont {S.~T.}\ \bibnamefont
  {Bramwell}}\ and\ \bibinfo {author} {\bibfnamefont {M.~J.}\ \bibnamefont
  {Gingras}},\ }\href {\doibase https://doi.org/10.1126/science.1064761}
  {\bibfield  {journal} {\bibinfo  {journal} {Science}\ }\textbf {\bibinfo
  {volume} {294}},\ \bibinfo {pages} {1495} (\bibinfo {year}
  {2001})}\BibitemShut {NoStop}%
\bibitem [{\citenamefont {Krizan}\ and\ \citenamefont
  {Cava}(2014)}]{Krizan2014}%
  \BibitemOpen
  \bibfield  {author} {\bibinfo {author} {\bibfnamefont {J.~W.}\ \bibnamefont
  {Krizan}}\ and\ \bibinfo {author} {\bibfnamefont {R.~J.}\ \bibnamefont
  {Cava}},\ }\href {\doibase 10.1103/PhysRevB.89.214401} {\bibfield  {journal}
  {\bibinfo  {journal} {Phys. Rev. B}\ }\textbf {\bibinfo {volume} {89}},\
  \bibinfo {pages} {214401} (\bibinfo {year} {2014})}\BibitemShut {NoStop}%
\bibitem [{\citenamefont {Krizan}\ and\ \citenamefont
  {Cava}(2015{\natexlab{a}})}]{krizan2015}%
  \BibitemOpen
  \bibfield  {author} {\bibinfo {author} {\bibfnamefont {J.~W.}\ \bibnamefont
  {Krizan}}\ and\ \bibinfo {author} {\bibfnamefont {R.~J.}\ \bibnamefont
  {Cava}},\ }\href {\doibase 10.1103/PhysRevB.92.014406} {\bibfield  {journal}
  {\bibinfo  {journal} {Phys. Rev. B}\ }\textbf {\bibinfo {volume} {92}},\
  \bibinfo {pages} {014406} (\bibinfo {year} {2015}{\natexlab{a}})}\BibitemShut
  {NoStop}%
\bibitem [{\citenamefont {Krizan}\ and\ \citenamefont
  {Cava}(2015{\natexlab{b}})}]{krizan2015nasrco2f7}%
  \BibitemOpen
  \bibfield  {author} {\bibinfo {author} {\bibfnamefont {J.}~\bibnamefont
  {Krizan}}\ and\ \bibinfo {author} {\bibfnamefont {R.~J.}\ \bibnamefont
  {Cava}},\ }\href {\doibase 10.1088/0953-8984/27/29/296002} {\bibfield
  {journal} {\bibinfo  {journal} {Journal of Physics: Condensed Matter}\
  }\textbf {\bibinfo {volume} {27}},\ \bibinfo {pages} {296002} (\bibinfo
  {year} {2015}{\natexlab{b}})}\BibitemShut {NoStop}%
\bibitem [{\citenamefont {Andreanov}\ \emph {et~al.}(2010)\citenamefont
  {Andreanov}, \citenamefont {Chalker}, \citenamefont {Saunders},\ and\
  \citenamefont {Sherrington}}]{Andreanov2010}%
  \BibitemOpen
  \bibfield  {author} {\bibinfo {author} {\bibfnamefont {A.}~\bibnamefont
  {Andreanov}}, \bibinfo {author} {\bibfnamefont {J.~T.}\ \bibnamefont
  {Chalker}}, \bibinfo {author} {\bibfnamefont {T.~E.}\ \bibnamefont
  {Saunders}}, \ and\ \bibinfo {author} {\bibfnamefont {D.}~\bibnamefont
  {Sherrington}},\ }\href {\doibase 10.1103/PhysRevB.81.014406} {\bibfield
  {journal} {\bibinfo  {journal} {Phys. Rev. B}\ }\textbf {\bibinfo {volume}
  {81}},\ \bibinfo {pages} {014406} (\bibinfo {year} {2010})}\BibitemShut
  {NoStop}%
\bibitem [{\citenamefont {Alexandradinata}\ \emph {et~al.}(2020)\citenamefont
  {Alexandradinata}, \citenamefont {Armitage}, \citenamefont {Baydin},
  \citenamefont {Bi}, \citenamefont {Cao}, \citenamefont {Changlani},
  \citenamefont {Chertkov}, \citenamefont {Neto}, \citenamefont {Delacretaz},
  \citenamefont {Baggari} \emph {et~al.}}]{alexandradinata2020future}%
  \BibitemOpen
  \bibfield  {author} {\bibinfo {author} {\bibfnamefont {A.}~\bibnamefont
  {Alexandradinata}}, \bibinfo {author} {\bibfnamefont {N.}~\bibnamefont
  {Armitage}}, \bibinfo {author} {\bibfnamefont {A.}~\bibnamefont {Baydin}},
  \bibinfo {author} {\bibfnamefont {W.}~\bibnamefont {Bi}}, \bibinfo {author}
  {\bibfnamefont {Y.}~\bibnamefont {Cao}}, \bibinfo {author} {\bibfnamefont
  {H.~J.}\ \bibnamefont {Changlani}}, \bibinfo {author} {\bibfnamefont
  {E.}~\bibnamefont {Chertkov}}, \bibinfo {author} {\bibfnamefont {E.~H.}\
  \bibnamefont {Neto}}, \bibinfo {author} {\bibfnamefont {L.}~\bibnamefont
  {Delacretaz}}, \bibinfo {author} {\bibfnamefont {I.~E.}\ \bibnamefont
  {Baggari}},  \emph {et~al.},\ }\href@noop {} {\bibfield  {journal} {\bibinfo
  {journal} {arXiv preprint arXiv:2010.00584}\ } (\bibinfo {year}
  {2020})}\BibitemShut {NoStop}%
\bibitem [{\citenamefont {Moessner}\ and\ \citenamefont
  {Chalker}(1998)}]{Moessner1998}%
  \BibitemOpen
  \bibfield  {author} {\bibinfo {author} {\bibfnamefont {R.}~\bibnamefont
  {Moessner}}\ and\ \bibinfo {author} {\bibfnamefont {J.~T.}\ \bibnamefont
  {Chalker}},\ }\href {\doibase 10.1103/PhysRevLett.80.2929} {\bibfield
  {journal} {\bibinfo  {journal} {Phys. Rev. Lett.}\ }\textbf {\bibinfo
  {volume} {80}},\ \bibinfo {pages} {2929} (\bibinfo {year}
  {1998})}\BibitemShut {NoStop}%
\bibitem [{\citenamefont {Yang}\ \emph {et~al.}(2015)\citenamefont {Yang},
  \citenamefont {Samarakoon}, \citenamefont {Dissanayake}, \citenamefont
  {Ueda}, \citenamefont {Klich}, \citenamefont {Iida}, \citenamefont
  {Pajerowski}, \citenamefont {Butch}, \citenamefont {Huang}, \citenamefont
  {Copley} \emph {et~al.}}]{yang2015spin}%
  \BibitemOpen
  \bibfield  {author} {\bibinfo {author} {\bibfnamefont {J.}~\bibnamefont
  {Yang}}, \bibinfo {author} {\bibfnamefont {A.}~\bibnamefont {Samarakoon}},
  \bibinfo {author} {\bibfnamefont {S.}~\bibnamefont {Dissanayake}}, \bibinfo
  {author} {\bibfnamefont {H.}~\bibnamefont {Ueda}}, \bibinfo {author}
  {\bibfnamefont {I.}~\bibnamefont {Klich}}, \bibinfo {author} {\bibfnamefont
  {K.}~\bibnamefont {Iida}}, \bibinfo {author} {\bibfnamefont {D.}~\bibnamefont
  {Pajerowski}}, \bibinfo {author} {\bibfnamefont {N.~P.}\ \bibnamefont
  {Butch}}, \bibinfo {author} {\bibfnamefont {Q.}~\bibnamefont {Huang}},
  \bibinfo {author} {\bibfnamefont {J.~R.}\ \bibnamefont {Copley}},  \emph
  {et~al.},\ }\href {\doibase 10.1073/pnas.1503126112} {\bibfield  {journal}
  {\bibinfo  {journal} {Proceedings of the National Academy of Sciences}\
  }\textbf {\bibinfo {volume} {112}},\ \bibinfo {pages} {11519} (\bibinfo
  {year} {2015})}\BibitemShut {NoStop}%
\bibitem [{\citenamefont {Cepas}\ and\ \citenamefont
  {Canals}(2012)}]{cepas2012heterogeneous}%
  \BibitemOpen
  \bibfield  {author} {\bibinfo {author} {\bibfnamefont {O.}~\bibnamefont
  {Cepas}}\ and\ \bibinfo {author} {\bibfnamefont {B.}~\bibnamefont {Canals}},\
  }\href {\doibase 10.1103/PhysRevB.86.024434} {\bibfield  {journal} {\bibinfo
  {journal} {Physical Review B}\ }\textbf {\bibinfo {volume} {86}},\ \bibinfo
  {pages} {024434} (\bibinfo {year} {2012})}\BibitemShut {NoStop}%
\bibitem [{\citenamefont {Moriya}(1960)}]{Moriya1960}%
  \BibitemOpen
  \bibfield  {author} {\bibinfo {author} {\bibfnamefont {T.}~\bibnamefont
  {Moriya}},\ }\href {\doibase 10.1103/PhysRev.120.91} {\bibfield  {journal}
  {\bibinfo  {journal} {Phys. Rev.}\ }\textbf {\bibinfo {volume} {120}},\
  \bibinfo {pages} {91} (\bibinfo {year} {1960})}\BibitemShut {NoStop}%
\bibitem [{\citenamefont {Elhajal}\ \emph {et~al.}(2005)\citenamefont
  {Elhajal}, \citenamefont {Canals}, \citenamefont {Sunyer},\ and\
  \citenamefont {Lacroix}}]{Elhajal2005}%
  \BibitemOpen
  \bibfield  {author} {\bibinfo {author} {\bibfnamefont {M.}~\bibnamefont
  {Elhajal}}, \bibinfo {author} {\bibfnamefont {B.}~\bibnamefont {Canals}},
  \bibinfo {author} {\bibfnamefont {R.}~\bibnamefont {Sunyer}}, \ and\ \bibinfo
  {author} {\bibfnamefont {C.}~\bibnamefont {Lacroix}},\ }\href {\doibase
  10.1103/PhysRevB.71.094420} {\bibfield  {journal} {\bibinfo  {journal} {Phys.
  Rev. B}\ }\textbf {\bibinfo {volume} {71}},\ \bibinfo {pages} {094420}
  (\bibinfo {year} {2005})}\BibitemShut {NoStop}%
\bibitem [{\citenamefont {Kennedy}(1997)}]{KENNEDY1997}%
  \BibitemOpen
  \bibfield  {author} {\bibinfo {author} {\bibfnamefont {B.~J.}\ \bibnamefont
  {Kennedy}},\ }\href {\doibase https://doi.org/10.1016/S0921-4526(97)00570-X}
  {\bibfield  {journal} {\bibinfo  {journal} {Physica B: Condensed Matter}\
  }\textbf {\bibinfo {volume} {241-243}},\ \bibinfo {pages} {303 } (\bibinfo
  {year} {1997})},\ \bibinfo {note} {proceedings of the International
  Conference on Neutron Scattering}\BibitemShut {NoStop}%
\bibitem [{\citenamefont {Wills}\ \emph {et~al.}(2006)\citenamefont {Wills},
  \citenamefont {Zhitomirsky}, \citenamefont {Canals}, \citenamefont {Sanchez},
  \citenamefont {Bonville}, \citenamefont {de~R{\'{e}}otier},\ and\
  \citenamefont {Yaouanc}}]{Wills2006}%
  \BibitemOpen
  \bibfield  {author} {\bibinfo {author} {\bibfnamefont {A.~S.}\ \bibnamefont
  {Wills}}, \bibinfo {author} {\bibfnamefont {M.~E.}\ \bibnamefont
  {Zhitomirsky}}, \bibinfo {author} {\bibfnamefont {B.}~\bibnamefont {Canals}},
  \bibinfo {author} {\bibfnamefont {J.~P.}\ \bibnamefont {Sanchez}}, \bibinfo
  {author} {\bibfnamefont {P.}~\bibnamefont {Bonville}}, \bibinfo {author}
  {\bibfnamefont {P.~D.}\ \bibnamefont {de~R{\'{e}}otier}}, \ and\ \bibinfo
  {author} {\bibfnamefont {A.}~\bibnamefont {Yaouanc}},\ }\href {\doibase
  10.1088/0953-8984/18/3/l02} {\bibfield  {journal} {\bibinfo  {journal}
  {Journal of Physics: Condensed Matter}\ }\textbf {\bibinfo {volume} {18}},\
  \bibinfo {pages} {L37} (\bibinfo {year} {2006})}\BibitemShut {NoStop}%
\bibitem [{\citenamefont {Sadeghi}\ \emph {et~al.}(2015)\citenamefont
  {Sadeghi}, \citenamefont {Alaei}, \citenamefont {Shahbazi},\ and\
  \citenamefont {Gingras}}]{Sadeghi2015}%
  \BibitemOpen
  \bibfield  {author} {\bibinfo {author} {\bibfnamefont {A.}~\bibnamefont
  {Sadeghi}}, \bibinfo {author} {\bibfnamefont {M.}~\bibnamefont {Alaei}},
  \bibinfo {author} {\bibfnamefont {F.}~\bibnamefont {Shahbazi}}, \ and\
  \bibinfo {author} {\bibfnamefont {M.~J.~P.}\ \bibnamefont {Gingras}},\ }\href
  {\doibase 10.1103/PhysRevB.91.140407} {\bibfield  {journal} {\bibinfo
  {journal} {Phys. Rev. B}\ }\textbf {\bibinfo {volume} {91}},\ \bibinfo
  {pages} {140407} (\bibinfo {year} {2015})}\BibitemShut {NoStop}%
\bibitem [{\citenamefont {Xiang}\ \emph {et~al.}(2011)\citenamefont {Xiang},
  \citenamefont {Kan}, \citenamefont {Whangbo}, \citenamefont {Lee},
  \citenamefont {Wei},\ and\ \citenamefont {Gong}}]{Xiang2011}%
  \BibitemOpen
  \bibfield  {author} {\bibinfo {author} {\bibfnamefont {H.~J.}\ \bibnamefont
  {Xiang}}, \bibinfo {author} {\bibfnamefont {E.~J.}\ \bibnamefont {Kan}},
  \bibinfo {author} {\bibfnamefont {M.-H.}\ \bibnamefont {Whangbo}}, \bibinfo
  {author} {\bibfnamefont {C.}~\bibnamefont {Lee}}, \bibinfo {author}
  {\bibfnamefont {S.-H.}\ \bibnamefont {Wei}}, \ and\ \bibinfo {author}
  {\bibfnamefont {X.~G.}\ \bibnamefont {Gong}},\ }\href@noop {} {\bibfield
  {journal} {\bibinfo  {journal} {Phys. Rev. B}\ }\textbf {\bibinfo {volume}
  {83}},\ \bibinfo {pages} {174402} (\bibinfo {year} {2011})}\BibitemShut
  {NoStop}%
\bibitem [{\citenamefont {FLEURgroup}()}]{fleur}%
  \BibitemOpen
  \bibfield  {author} {\bibinfo {author} {\bibnamefont {FLEURgroup}},\ }\href
  {http://www.flapw.de/} {\enquote {\bibinfo {title} {http://www.flapw.de/},}\
  }\BibitemShut {NoStop}%
  \bibitem [{\citenamefont {Giannozzi}\ \emph {et~al.}(2009)\citenamefont
  {Giannozzi}, \citenamefont {Baroni}, \citenamefont {Bonini}, \citenamefont
  {Calandra}, \citenamefont {Car}, \citenamefont {Cavazzoni}, \citenamefont
  {Ceresoli}, \citenamefont {Chiarotti}, \citenamefont {Cococcioni},
  \citenamefont {Dabo}, \citenamefont {Corso}, \citenamefont {de~Gironcoli},
  \citenamefont {Fabris}, \citenamefont {Fratesi}, \citenamefont {Gebauer},
  \citenamefont {Gerstmann}, \citenamefont {Gougoussis}, \citenamefont
  {Kokalj}, \citenamefont {Lazzeri}, \citenamefont {Martin-Samos},
  \citenamefont {Marzari}, \citenamefont {Mauri}, \citenamefont {Mazzarello},
  \citenamefont {Paolini}, \citenamefont {Pasquarello}, \citenamefont
  {Paulatto}, \citenamefont {Sbraccia}, \citenamefont {Scandolo}, \citenamefont
  {Sclauzero}, \citenamefont {Seitsonen}, \citenamefont {Smogunov},
  \citenamefont {Umari},\ and\ \citenamefont {Wentzcovitch}}]{Giannozzi_2009}%
  \BibitemOpen
  \bibfield  {author} {\bibinfo {author} {\bibfnamefont {P.}~\bibnamefont
  {Giannozzi}}, \bibinfo {author} {\bibfnamefont {S.}~\bibnamefont {Baroni}},
  \bibinfo {author} {\bibfnamefont {N.}~\bibnamefont {Bonini}}, \bibinfo
  {author} {\bibfnamefont {M.}~\bibnamefont {Calandra}}, \bibinfo {author}
  {\bibfnamefont {R.}~\bibnamefont {Car}}, \bibinfo {author} {\bibfnamefont
  {C.}~\bibnamefont {Cavazzoni}}, \bibinfo {author} {\bibfnamefont
  {D.}~\bibnamefont {Ceresoli}}, \bibinfo {author} {\bibfnamefont {G.~L.}\
  \bibnamefont {Chiarotti}}, \bibinfo {author} {\bibfnamefont {M.}~\bibnamefont
  {Cococcioni}}, \bibinfo {author} {\bibfnamefont {I.}~\bibnamefont {Dabo}},
  \bibinfo {author} {\bibfnamefont {A.~D.}\ \bibnamefont {Corso}}, \bibinfo
  {author} {\bibfnamefont {S.}~\bibnamefont {de~Gironcoli}}, \bibinfo {author}
  {\bibfnamefont {S.}~\bibnamefont {Fabris}}, \bibinfo {author} {\bibfnamefont
  {G.}~\bibnamefont {Fratesi}}, \bibinfo {author} {\bibfnamefont
  {R.}~\bibnamefont {Gebauer}}, \bibinfo {author} {\bibfnamefont
  {U.}~\bibnamefont {Gerstmann}}, \bibinfo {author} {\bibfnamefont
  {C.}~\bibnamefont {Gougoussis}}, \bibinfo {author} {\bibfnamefont
  {A.}~\bibnamefont {Kokalj}}, \bibinfo {author} {\bibfnamefont
  {M.}~\bibnamefont {Lazzeri}}, \bibinfo {author} {\bibfnamefont
  {L.}~\bibnamefont {Martin-Samos}}, \bibinfo {author} {\bibfnamefont
  {N.}~\bibnamefont {Marzari}}, \bibinfo {author} {\bibfnamefont
  {F.}~\bibnamefont {Mauri}}, \bibinfo {author} {\bibfnamefont
  {R.}~\bibnamefont {Mazzarello}}, \bibinfo {author} {\bibfnamefont
  {S.}~\bibnamefont {Paolini}}, \bibinfo {author} {\bibfnamefont
  {A.}~\bibnamefont {Pasquarello}}, \bibinfo {author} {\bibfnamefont
  {L.}~\bibnamefont {Paulatto}}, \bibinfo {author} {\bibfnamefont
  {C.}~\bibnamefont {Sbraccia}}, \bibinfo {author} {\bibfnamefont
  {S.}~\bibnamefont {Scandolo}}, \bibinfo {author} {\bibfnamefont
  {G.}~\bibnamefont {Sclauzero}}, \bibinfo {author} {\bibfnamefont {A.~P.}\
  \bibnamefont {Seitsonen}}, \bibinfo {author} {\bibfnamefont {A.}~\bibnamefont
  {Smogunov}}, \bibinfo {author} {\bibfnamefont {P.}~\bibnamefont {Umari}}, \
  and\ \bibinfo {author} {\bibfnamefont {R.~M.}\ \bibnamefont {Wentzcovitch}},\
  }\href {\doibase 10.1088/0953-8984/21/39/395502} {\bibfield  {journal}
  {\bibinfo  {journal} {Journal of Physics: Condensed Matter}\ }\textbf
  {\bibinfo {volume} {21}},\ \bibinfo {pages} {395502} (\bibinfo {year}
  {2009})}\BibitemShut {NoStop}%
\bibitem [{\citenamefont {Giannozzi}\ \emph {et~al.}(2017)\citenamefont
  {Giannozzi}, \citenamefont {Andreussi}, \citenamefont {Brumme}, \citenamefont
  {Bunau}, \citenamefont {Nardelli}, \citenamefont {Calandra}, \citenamefont
  {Car}, \citenamefont {Cavazzoni}, \citenamefont {Ceresoli}, \citenamefont
  {Cococcioni}, \citenamefont {Colonna}, \citenamefont {Carnimeo},
  \citenamefont {Corso}, \citenamefont {de~Gironcoli}, \citenamefont {Delugas},
  \citenamefont {DiStasio}, \citenamefont {Ferretti}, \citenamefont {Floris},
  \citenamefont {Fratesi}, \citenamefont {Fugallo}, \citenamefont {Gebauer},
  \citenamefont {Gerstmann}, \citenamefont {Giustino}, \citenamefont {Gorni},
  \citenamefont {Jia}, \citenamefont {Kawamura}, \citenamefont {Ko},
  \citenamefont {Kokalj}, \citenamefont {Kü{\c{c}}ükbenli}, \citenamefont
  {Lazzeri}, \citenamefont {Marsili}, \citenamefont {Marzari}, \citenamefont
  {Mauri}, \citenamefont {Nguyen}, \citenamefont {Nguyen}, \citenamefont {de-la
  Roza}, \citenamefont {Paulatto}, \citenamefont {Ponc{\'{e}}}, \citenamefont
  {Rocca}, \citenamefont {Sabatini}, \citenamefont {Santra}, \citenamefont
  {Schlipf}, \citenamefont {Seitsonen}, \citenamefont {Smogunov}, \citenamefont
  {Timrov}, \citenamefont {Thonhauser}, \citenamefont {Umari}, \citenamefont
  {Vast}, \citenamefont {Wu},\ and\ \citenamefont {Baroni}}]{Giannozzi_2017}%
  \BibitemOpen
  \bibfield  {author} {\bibinfo {author} {\bibfnamefont {P.}~\bibnamefont
  {Giannozzi}}, \bibinfo {author} {\bibfnamefont {O.}~\bibnamefont
  {Andreussi}}, \bibinfo {author} {\bibfnamefont {T.}~\bibnamefont {Brumme}},
  \bibinfo {author} {\bibfnamefont {O.}~\bibnamefont {Bunau}}, \bibinfo
  {author} {\bibfnamefont {M.~B.}\ \bibnamefont {Nardelli}}, \bibinfo {author}
  {\bibfnamefont {M.}~\bibnamefont {Calandra}}, \bibinfo {author}
  {\bibfnamefont {R.}~\bibnamefont {Car}}, \bibinfo {author} {\bibfnamefont
  {C.}~\bibnamefont {Cavazzoni}}, \bibinfo {author} {\bibfnamefont
  {D.}~\bibnamefont {Ceresoli}}, \bibinfo {author} {\bibfnamefont
  {M.}~\bibnamefont {Cococcioni}}, \bibinfo {author} {\bibfnamefont
  {N.}~\bibnamefont {Colonna}}, \bibinfo {author} {\bibfnamefont
  {I.}~\bibnamefont {Carnimeo}}, \bibinfo {author} {\bibfnamefont {A.~D.}\
  \bibnamefont {Corso}}, \bibinfo {author} {\bibfnamefont {S.}~\bibnamefont
  {de~Gironcoli}}, \bibinfo {author} {\bibfnamefont {P.}~\bibnamefont
  {Delugas}}, \bibinfo {author} {\bibfnamefont {R.~A.}\ \bibnamefont
  {DiStasio}}, \bibinfo {author} {\bibfnamefont {A.}~\bibnamefont {Ferretti}},
  \bibinfo {author} {\bibfnamefont {A.}~\bibnamefont {Floris}}, \bibinfo
  {author} {\bibfnamefont {G.}~\bibnamefont {Fratesi}}, \bibinfo {author}
  {\bibfnamefont {G.}~\bibnamefont {Fugallo}}, \bibinfo {author} {\bibfnamefont
  {R.}~\bibnamefont {Gebauer}}, \bibinfo {author} {\bibfnamefont
  {U.}~\bibnamefont {Gerstmann}}, \bibinfo {author} {\bibfnamefont
  {F.}~\bibnamefont {Giustino}}, \bibinfo {author} {\bibfnamefont
  {T.}~\bibnamefont {Gorni}}, \bibinfo {author} {\bibfnamefont
  {J.}~\bibnamefont {Jia}}, \bibinfo {author} {\bibfnamefont {M.}~\bibnamefont
  {Kawamura}}, \bibinfo {author} {\bibfnamefont {H.-Y.}\ \bibnamefont {Ko}},
  \bibinfo {author} {\bibfnamefont {A.}~\bibnamefont {Kokalj}}, \bibinfo
  {author} {\bibfnamefont {E.}~\bibnamefont {Kü{\c{c}}ükbenli}}, \bibinfo
  {author} {\bibfnamefont {M.}~\bibnamefont {Lazzeri}}, \bibinfo {author}
  {\bibfnamefont {M.}~\bibnamefont {Marsili}}, \bibinfo {author} {\bibfnamefont
  {N.}~\bibnamefont {Marzari}}, \bibinfo {author} {\bibfnamefont
  {F.}~\bibnamefont {Mauri}}, \bibinfo {author} {\bibfnamefont {N.~L.}\
  \bibnamefont {Nguyen}}, \bibinfo {author} {\bibfnamefont {H.-V.}\
  \bibnamefont {Nguyen}}, \bibinfo {author} {\bibfnamefont {A.~O.}\
  \bibnamefont {de-la Roza}}, \bibinfo {author} {\bibfnamefont
  {L.}~\bibnamefont {Paulatto}}, \bibinfo {author} {\bibfnamefont
  {S.}~\bibnamefont {Ponc{\'{e}}}}, \bibinfo {author} {\bibfnamefont
  {D.}~\bibnamefont {Rocca}}, \bibinfo {author} {\bibfnamefont
  {R.}~\bibnamefont {Sabatini}}, \bibinfo {author} {\bibfnamefont
  {B.}~\bibnamefont {Santra}}, \bibinfo {author} {\bibfnamefont
  {M.}~\bibnamefont {Schlipf}}, \bibinfo {author} {\bibfnamefont {A.~P.}\
  \bibnamefont {Seitsonen}}, \bibinfo {author} {\bibfnamefont {A.}~\bibnamefont
  {Smogunov}}, \bibinfo {author} {\bibfnamefont {I.}~\bibnamefont {Timrov}},
  \bibinfo {author} {\bibfnamefont {T.}~\bibnamefont {Thonhauser}}, \bibinfo
  {author} {\bibfnamefont {P.}~\bibnamefont {Umari}}, \bibinfo {author}
  {\bibfnamefont {N.}~\bibnamefont {Vast}}, \bibinfo {author} {\bibfnamefont
  {X.}~\bibnamefont {Wu}}, \ and\ \bibinfo {author} {\bibfnamefont
  {S.}~\bibnamefont {Baroni}},\ }\href {\doibase 10.1088/1361-648x/aa8f79}
  {\bibfield  {journal} {\bibinfo  {journal} {Journal of Physics: Condensed
  Matter}\ }\textbf {\bibinfo {volume} {29}},\ \bibinfo {pages} {465901}
  (\bibinfo {year} {2017})}\BibitemShut {NoStop}%
\bibitem [{\citenamefont {Perdew}\ \emph {et~al.}(1996)\citenamefont {Perdew},
  \citenamefont {Burke},\ and\ \citenamefont {Ernzerhof}}]{Perdew1996}%
  \BibitemOpen
  \bibfield  {author} {\bibinfo {author} {\bibfnamefont {J.~P.}\ \bibnamefont
  {Perdew}}, \bibinfo {author} {\bibfnamefont {K.}~\bibnamefont {Burke}}, \
  and\ \bibinfo {author} {\bibfnamefont {M.}~\bibnamefont {Ernzerhof}},\ }\href
  {\doibase 10.1103/PhysRevLett.77.3865} {\bibfield  {journal} {\bibinfo
  {journal} {Phys. Rev. Lett.}\ }\textbf {\bibinfo {volume} {77}},\ \bibinfo
  {pages} {3865} (\bibinfo {year} {1996})}\BibitemShut {NoStop}%
\bibitem [{\citenamefont {Cococcioni}\ and\ \citenamefont
  {de~Gironcoli}(2005)}]{Cococcioni2005}%
  \BibitemOpen
  \bibfield  {author} {\bibinfo {author} {\bibfnamefont {M.}~\bibnamefont
  {Cococcioni}}\ and\ \bibinfo {author} {\bibfnamefont {S.}~\bibnamefont
  {de~Gironcoli}},\ }\href {\doibase 10.1103/PhysRevB.71.035105} {\bibfield
  {journal} {\bibinfo  {journal} {Phys. Rev. B}\ }\textbf {\bibinfo {volume}
  {71}},\ \bibinfo {pages} {035105} (\bibinfo {year} {2005})}\BibitemShut
  {NoStop}%
\bibitem [{\citenamefont {Timrov}\ \emph {et~al.}(2018)\citenamefont {Timrov},
  \citenamefont {Marzari},\ and\ \citenamefont {Cococcioni}}]{Timrov2018}%
  \BibitemOpen
  \bibfield  {author} {\bibinfo {author} {\bibfnamefont {I.}~\bibnamefont
  {Timrov}}, \bibinfo {author} {\bibfnamefont {N.}~\bibnamefont {Marzari}}, \
  and\ \bibinfo {author} {\bibfnamefont {M.}~\bibnamefont {Cococcioni}},\
  }\href {\doibase 10.1103/PhysRevB.98.085127} {\bibfield  {journal} {\bibinfo
  {journal} {Phys. Rev. B}\ }\textbf {\bibinfo {volume} {98}},\ \bibinfo
  {pages} {085127} (\bibinfo {year} {2018})}\BibitemShut {NoStop}%
\bibitem [{\citenamefont {Garrity}\ \emph {et~al.}(2014)\citenamefont
  {Garrity}, \citenamefont {Bennett}, \citenamefont {Rabe},\ and\ \citenamefont
  {Vanderbilt}}]{Vanderbilt2014}%
  \BibitemOpen
  \bibfield  {author} {\bibinfo {author} {\bibfnamefont {K.~F.}\ \bibnamefont
  {Garrity}}, \bibinfo {author} {\bibfnamefont {J.~W.}\ \bibnamefont
  {Bennett}}, \bibinfo {author} {\bibfnamefont {K.~M.}\ \bibnamefont {Rabe}}, \
  and\ \bibinfo {author} {\bibfnamefont {D.}~\bibnamefont {Vanderbilt}},\
  }\href {\doibase https://doi.org/10.1016/j.commatsci.2013.08.053} {\bibfield
  {journal} {\bibinfo  {journal} {Computational Materials Science}\ }\textbf
  {\bibinfo {volume} {81}},\ \bibinfo {pages} {446 } (\bibinfo {year}
  {2014})}\BibitemShut {NoStop}%  
  \bibitem [{\citenamefont {Okhotnikov}\ \emph {et~al.}(2016)\citenamefont
  {Okhotnikov}, \citenamefont {Charpentier},\ and\ \citenamefont
  {Cadars}}]{Okhotnikov2016}%
  \BibitemOpen
  \bibfield  {author} {\bibinfo {author} {\bibfnamefont {K.}~\bibnamefont
  {Okhotnikov}}, \bibinfo {author} {\bibfnamefont {T.}~\bibnamefont
  {Charpentier}}, \ and\ \bibinfo {author} {\bibfnamefont {S.}~\bibnamefont
  {Cadars}},\ }\href {\doibase 10.1186/s13321-016-0129-3} {\bibfield  {journal}
  {\bibinfo  {journal} {Journal of Cheminformatics}\ }\textbf {\bibinfo
  {volume} {8}},\ \bibinfo {pages} {17} (\bibinfo {year} {2016})}\BibitemShut
  {NoStop}%
\bibitem [{\citenamefont {Hukushima}\ and\ \citenamefont
  {Nemoto}(1996)}]{Hukushima1996}%
  \BibitemOpen
  \bibfield  {author} {\bibinfo {author} {\bibfnamefont {K.}~\bibnamefont
  {Hukushima}}\ and\ \bibinfo {author} {\bibfnamefont {K.}~\bibnamefont
  {Nemoto}},\ }\href {\doibase 10.1143/JPSJ.65.1604} {\bibfield  {journal}
  {\bibinfo  {journal} {Journal of the Physical Society of Japan}\ }\textbf
  {\bibinfo {volume} {65}},\ \bibinfo {pages} {1604} (\bibinfo {year}
  {1996})}\BibitemShut {NoStop}%
\bibitem [{\citenamefont {Rezaei}\ \emph {et~al.}(2022)\citenamefont {Rezaei},
  \citenamefont {Alaei},\ and\ \citenamefont {Akbarzadeh}}]{REZAEI2022110947}%
  \BibitemOpen
  \bibfield  {author} {\bibinfo {author} {\bibfnamefont {N.}~\bibnamefont
  {Rezaei}}, \bibinfo {author} {\bibfnamefont {M.}~\bibnamefont {Alaei}}, \
  and\ \bibinfo {author} {\bibfnamefont {H.}~\bibnamefont {Akbarzadeh}},\
  }\href {\doibase https://doi.org/10.1016/j.commatsci.2021.110947} {\bibfield
  {journal} {\bibinfo  {journal} {Computational Materials Science}\ }\textbf
  {\bibinfo {volume} {202}},\ \bibinfo {pages} {110947} (\bibinfo {year}
  {2022})}\BibitemShut {NoStop}%
\bibitem [{\citenamefont {Coey}\ \emph {et~al.}(2013)\citenamefont {Coey},
  \citenamefont {Venkatesan},\ and\ \citenamefont {Xu}}]{coey2013introduction}%
  \BibitemOpen
  \bibfield  {author} {\bibinfo {author} {\bibfnamefont {J.~M.~D.}\
  \bibnamefont {Coey}}, \bibinfo {author} {\bibfnamefont {M.}~\bibnamefont
  {Venkatesan}}, \ and\ \bibinfo {author} {\bibfnamefont {H.}~\bibnamefont
  {Xu}},\ }\enquote {\bibinfo {title} {Introduction to magnetic oxides},}\ in\
  \href {\doibase https://doi.org/10.1002/9783527654864.ch1} {\emph {\bibinfo
  {booktitle} {Functional Metal Oxides}}}\ (\bibinfo  {publisher} {John Wiley
  and Sons, Ltd},\ \bibinfo {year} {2013})\ Chap.~\bibinfo {chapter} {1}, pp.\
  \bibinfo {pages} {1--49}\BibitemShut {NoStop}%
\bibitem [{\citenamefont {Vaugier}\ \emph {et~al.}(2012)\citenamefont
  {Vaugier}, \citenamefont {Jiang},\ and\ \citenamefont
  {Biermann}}]{Vaugier2012}%
  \BibitemOpen
  \bibfield  {author} {\bibinfo {author} {\bibfnamefont {L.}~\bibnamefont
  {Vaugier}}, \bibinfo {author} {\bibfnamefont {H.}~\bibnamefont {Jiang}}, \
  and\ \bibinfo {author} {\bibfnamefont {S.}~\bibnamefont {Biermann}},\ }\href
  {\doibase 10.1103/PhysRevB.86.165105} {\bibfield  {journal} {\bibinfo
  {journal} {Phys. Rev. B}\ }\textbf {\bibinfo {volume} {86}},\ \bibinfo
  {pages} {165105} (\bibinfo {year} {2012})}\BibitemShut {NoStop}%
\bibitem [{\citenamefont {Saunders}\ and\ \citenamefont
  {Chalker}(2007)}]{saunders2007spin}%
  \BibitemOpen
  \bibfield  {author} {\bibinfo {author} {\bibfnamefont {T.}~\bibnamefont
  {Saunders}}\ and\ \bibinfo {author} {\bibfnamefont {J.}~\bibnamefont
  {Chalker}},\ }\href {\doibase 10.1103/PhysRevLett.98.157201} {\bibfield
  {journal} {\bibinfo  {journal} {Physical review letters}\ }\textbf {\bibinfo
  {volume} {98}},\ \bibinfo {pages} {157201} (\bibinfo {year}
  {2007})}\BibitemShut {NoStop}%
\end{thebibliography}
\end{document}